\documentclass[letterpaper,11pt]{article}
\pdfoutput=1
\usepackage{jheppub}
\usepackage{slashed}
\usepackage{graphicx}
\usepackage{subfigure}
\usepackage{soul}
\usepackage{amsmath}
\usepackage{multirow}
\usepackage{tablefootnote}
\usepackage{hyperref}
\usepackage{xcolor}
\usepackage{epstopdf}
\usepackage{lscape}% get a Horizontal version
\usepackage{url}% \url{website}

\usepackage{xcolor}
\usepackage{float}
\usepackage[utf8]{inputenc}
\usepackage{feynmf}
\usepackage[compat=1.0.0]{tikz-feynman}

\newcommand{\beq}{\begin{equation}}
\newcommand{\eeq}{\end{equation}}
\newcommand{\bea}{\begin{eqnarray}}
\newcommand{\eea}{\end{eqnarray}}

\def\m1{M_1}
\def\m2{M_2}
\def\m3{M_3}

\def\ch10{\tilde \chi^0_1}

\newcommand{\lsim}{\mathrel{\mathop{\kern 0pt \rlap
  {\raise.2ex\hbox{$<$}}}
  \lower.9ex\hbox{\kern-.190em $\sim$}}}
\newcommand{\gsim}{\mathrel{\mathop{\kern 0pt \rlap
  {\raise.2ex\hbox{$>$}}}
  \lower.9ex\hbox{\kern-.190em $\sim$}}}

\definecolor{pink}{RGB}{255,105,180}

\definecolor{green2}{rgb}{0,0.56,0.32}

\def\figureautorefname~#1\null{Fig.\,#1\null}
\def\tableautorefname~#1\null{Tab.\,#1\null}

\def\equationautorefname~#1\null{Eq.\,(#1)\null}

\title{The electroweak precision constraints of the 2HDM+S}

\author[a]{Cheng Li}
\author[a]{, Juxiang Li}
\author[b]{, Shufang Su}
\author[a,c]{ and Wei Su}

\affiliation[a]{School of Science, Shenzhen Campus of Sun Yat-sen University, No. 66, Gongchang Road, \\ Guangming District, Shenzhen, Guangdong 518107, China }
\affiliation[b]{Department of Physics, University of Arizona, Tucson, AZ 85721, U.S.A.}
\affiliation[c]{Institute of Theoretical Physics, Chinese Academy of Sciences, Beijing 100190, P. R. China}

\emailAdd{lich389@mail.sysu.edu.cn}
\emailAdd{lijx376@mail2.sysu.edu.cn}
\emailAdd{suwei26@mail.sysu.edu.cn}
\emailAdd{shufang@email.arizona.edu}

\abstract{The 2HDM+S is the singlet extension of the Two-Higgs-Doublets Model (2HDM). The singlet field and its mixing with the 2HDM Higgs sector lead to new contributions to the electroweak precision observables, in particular, the oblique parameters.  In this paper, we identify five benchmark cases, where at most one mixing angle is nonzero and analyze the 95\% C.L. allowed parameter space by the oblique parameters. In the alignment limit of the 2HDM, we find that other than the usual mass relations of $m_H\sim m_{H^\pm}$ or $m_A\sim m_{H^\pm}$, electroweak precision measurements also impose an upper limit on the neutral Higgs masses. In the cases with nonzero singlet mixing with the 2HDM Higgses $H$ or $A$, we find approximate mass relations of $c^2_{\alpha_{HS}} m_{H} + s^2_{\alpha_{HS}}m_{h_S} = m_{H^\pm}$ or $c^2_{\alpha_{AS}} m_{A} + s^2_{\alpha_{AS}}m_{A_S} = m_{H^\pm}$. Those relations are universal to the 2HDM+S models, with or  without further symmetry assumption.  We also study the non-alignment limit of the 2HDM+S, which typically has tighter constraints on the masses and mixing angles.  At the end, we examine the complementarity between the electroweak precision analyses and the Higgs coupling precision measurements. }

\begin{document}
\maketitle
\flushbottom
%%%%%%%%%%%%%%%%%%%%%%%%%%%%%%%%%%%%%%%%%%%%%%
%%%%%%%%%%%%%% section %%%%%%%%%%%%%%%%%%%%%%%
\section{Introduction}
\label{sec:intro}

Electroweak precision observables have provided a precise test of the Standard Model (SM) at the loop level~\cite{Salam:2022izo,Gu:2017ckc}, which are consistent with the observations of a 125 GeV SM-like Higgs~\cite{ATLAS:2012yve,CMS:2012qbp}.  However, SM could not provide satisfactory solutions to dark matter, neutrino mass, baryogenesis, etc.\cite{Crivellin:2023zui,Georgi:1974yf,Cohen:1993nk,Bertone:2004pz}.  Furthermore, the naturalness problem in the SM also points to new physics beyond the SM~\cite{Martin:1997ns}. 

One of the simplest extensions of the SM Higgs sector is the Two-Higgs doublet model (2HDM)~\cite{Branco:2011iw}, which has been studied extensively in the literature. The 2HDM can be further extended by an additional singlet field, which is the N2HDM with a real singlet \cite{Muhlleitner:2016mzt} and the 2HDM+S with a complex singlet~\cite{Baum:2018zhf}. The 2HDM+S matches to the Next-to Minimal Supersymmetric Standard Model (NMSSM)~\cite{Ellwanger:2009dp} at low energy scale, and can provide a dark matter candidate~\cite{Dutta:2023cig}, as well as accommodate the possible 95~GeV excess at the LEP and the LHC~\cite{Heinemeyer:2021msz}. The phenomenological properties of 2HDM+S have only been explored in some specific scenarios, while the more general cases of 2HDM+S have not yet been studied in detail. In this work, we explore the implications of the electroweak precision measurements on the 2HDM+S parameter space. In particular, we focus on the oblique parameters $S$, $T$ and $U$, which are sensitive to the new physics contributions to the $W$ and $Z$ self-energies~\cite{Peskin:1991sw}.

The scalar sector of 2HDM + S includes two ${\rm SU(2)}_L$ doublets and a complex singlet. 
 The singlet field does not couple to the SM gauge bosons and fermions.  After the neutral components get vacuum expectation values (vev), assuming no CP-violation,  the mass spectrum of the Higgs sectors includes 3 CP-even scalars, two CP-odd scalars, and a pair of charged Higgses. In particular, the CP-even and CP-odd singlet components mix with the corresponding ones in the ${\rm SU(2)}_L$ doublets, which leads to the couplings of the singlet-like scalars to the SM gauge bosons, as well as modifies the couplings of the doublet-like scalars to the SM sector. The most general 2HDM+S Higgs potential has 27 free parameters, and eleven of those can be chosen to be the masses of the Higgs bosons, as well as the mixing angles between Higgses. The remaining parameters in the Higgs potential are the Higgs self-couplings, which do not directly contribute to the oblique parameters. Therefore, in our study, we only focus on those eleven mass and mixing parameters. We parametrize such mixing parameters by $\alpha_{hS}$, the mixing of the CP-even singlet with the 125 GeV SM-like Higgs $h$, 
  $\alpha_{HS}$, the mixing of the CP-even singlet with the 2HDM CP-even Higgs $H$, and $\alpha_{AS}$, the mixing of the CP-odd singlet with the 2HDM CP-odd Higgs $A$. Including the usual 2HDM mixing angle of the CP-even Higgses $\alpha$, we introduce five basic benchmark scenarios, Case-0 for the 2HDM alignment limit and Case-I $-$ IV in which only one mixing angle is set to be nonzero.  We analyze the contributions to the oblique parameters in each case and study the 95\% C.L. allowed region in the relevant parameter spaces under the oblique parameters.  After the discussion of these five benchmark scenarios, we also discuss the cases with a non-zero singlet mixing angle away from the alignment limit. 

The implications of electroweak precision measurements in the 2HDM have been studied in the literature~\cite{Grimus:2008nb, Haller:2018nnx}.  Our study offers a comprehensive electroweak precision analysis of 2HDM+S and identifies the impact of each singlet mixing angle. Since only the couplings between the Higgses and the SM gauge bosons enter the oblique parameters, our results are universal to the 2HDM+S models, with or  without further symmetry assumption of the Higgs potential. In addition, we  explore the complementarity of the electroweak precision analyses with the Higgs precision measurements.

The remainder of the paper is organized as follows. In \autoref{sec:thy}, we introduce the theoretical framework of 2HDM + S, as well as five benchmark cases. In \autoref{sec:EWP}, we introduce the electroweak oblique parameters and the contributions from the Higgs sector in the 2HDM+S. In \autoref{sec:res}, we present 95\% C.L. $STU$ allowed regions in the 2HDM+S parameter spaces of the five benchmark cases. In \autoref{sec:res2}, we study the cases beyond the alignment limit.  In \autoref{sec:Hprecision}, we show the complementarity of electroweak precision analyses with Higgs precision measurements. We conclude in \autoref{sec:conclu}.

% %%%%%%%%%%%%%%%%%%%%%%%%%%%%%%%%%%%%%%%%%%%%%%
%%%%%%%%%%%%%%%%%%%%%%%%%%%%%%%%%%%%%%%%%%%%%%
%%%%%%%%%%%%%% section %%%%%%%%%%%%%%%%%%%%%%%
\section{Theoretical framework}
\label{sec:thy}
The 2HDM+S is the singlet extension of the 2HDM, which has the following scalar contents:
\begin{equation}
\Phi_1=\begin{pmatrix}
\chi_1^+\\ \frac{{\rho_1+i\eta_1}}{\sqrt{2}}
\end{pmatrix},\qquad\Phi_2=\begin{pmatrix}
\chi_2^+\\ \frac{{\rho_2+i\eta_2}}{\sqrt{2}}
\end{pmatrix},\qquad S= \frac{\rho_S + i\eta_S}{\sqrt{2}},
\end{equation}
% \Shufang{Should we keep the neutral component convention the same, both with 1/sqrt(2)?}
where $\Phi_1$, $\Phi_2$ are the SU(2)$_L$  doublets with hypercharge $Y=1/2$, and $S$ is the gauge singlet. The general Higgs potential of 2HDM+S has been introduced in \cite{Baum:2018zhf}, while the simplified version of the 2HDM+S potential can be found in \cite{Heinemeyer:2021msz} when certain symmetries are imposed. After electroweak symmetry breaking, the neutral components of $\Phi_1$, $\Phi_2$ and $S$ develop  non-zero vacuum expectation values, $v_1$, $v_2$ and $v_S$, with $\sqrt{v_1^2 + v_2^2} = v\approx 246$~GeV.  We also introduce $\tan\beta = \frac{v_2}{v_1}$ with $\beta \in (0,\pi/2)$. Assuming no CP-violation, the mass spectrum of the 2HDM+S includes three neutral CP-even scalars, two neutral CP-odd scalars,  and one pair of charged Higgs bosons. 

The neutral CP-even states, $\rho_{1,2,S}$ mix together to form three mass eigenstates: the Non-SM-like $H$, the SM-like Higgs $h$ and the singlet-like $h_S$, with the $3\times3$ rotation matrix $R$ 
\begin{equation}
\begin{pmatrix}
        H\\ h\\ h_S
    \end{pmatrix} = R    \begin{pmatrix}
        \rho_1\\ \rho_2\\ \rho_S
    \end{pmatrix},\qquad R M_S^2 R^T = \operatorname{diag}\{m_H^2, m_{h}^2, m_{h_S}^2\}.
    \label{eq:roteven2}
    \end{equation}
the $R$ matrix is parametrized by three mixing angles $\alpha$, $\alpha_{HS}$, and $\alpha_{hS}$, which characterize the mixing angle between the two neutral components of the Higgs doublets $\rho_{1,2}$, and the mixing angles between the singlet $\rho_S$ with the 2HDM CP-even Higgses:
\begin{equation}
% \small
\begin{split}
  	R&= \begin{pmatrix}
       1& 0& 0\\
  	    0& c_{\alpha_{hS}}& s_{\alpha_{hS}}\\
       0& -s_{\alpha_{hS}}& c_{\alpha_{hS}}\\
  	\end{pmatrix}\begin{pmatrix}
  	    c_{\alpha_{HS}}& 0& s_{\alpha_{HS}}\\
       0& 1& 0\\
       -s_{\alpha_{HS}}& 0& c_{\alpha_{HS}}\\
       \end{pmatrix} \begin{pmatrix}
  	    c_{\alpha_{}}& s_{\alpha_{}}& 0\\
       -s_{\alpha_{}}& c_{\alpha_{}}& 0\\
       0& 0& 1
  	\end{pmatrix}\\
   &=\begin{pmatrix}
		c_{\alpha_{}}c_{\alpha_{HS}}& s_{\alpha_{}}c_{\alpha_{HS}}& s_{\alpha_{HS}}\\
		-s_{\alpha_{}}c_{\alpha_{hS}}-c_{\alpha_{}}s_{\alpha_{HS}}s_{\alpha_{hS}}& c_{\alpha_{}}c_{\alpha_{hS}}-s_{\alpha_{}}s_{\alpha_{HS}}s_{\alpha_{hS}}& c_{\alpha_{HS}}s_{\alpha_{hS}}\\
		s_{\alpha_{}}s_{\alpha_{hS}}-c_{\alpha_{}}s_{\alpha_{HS}}c_{\alpha_{hS}}& -s_{\alpha_{}}s_{\alpha_{HS}}c_{\alpha_{hS}}-c_{\alpha_{}}{{s_{\alpha_{hS}}}}& c_{\alpha_{HS}}c_{\alpha_{hS}}
	\end{pmatrix},
 \label{eq:roteven}
\end{split}
\end{equation} 

where we use the shorthand notation $s_{x} = \sin x$ and $c_{x} = \cos x$.
For the CP-odd states, we have 
\begin{equation}
\begin{pmatrix}
        G^0\\ A \\ A_S
    \end{pmatrix}=       \begin{pmatrix}
        1& ~0& ~0\\
        0& \multicolumn{2}{c}{\multirow{2}{*}{$R^A$}}& \\
        0& &  \\
    \end{pmatrix}
    \begin{pmatrix}
                c_{\beta}& s_{\beta}& 0\\
        -s_{\beta}& c_{\beta}& 0\\
        0 & 0& 1
    \end{pmatrix}\begin{pmatrix}
        \eta_1 \\ \eta_2 \\ \eta_S
    \end{pmatrix},\qquad
    R^A = \begin{pmatrix}
        c_{\alpha_{AS}}& s_{\alpha_{AS}}\\
        -s_{\alpha_{AS}}& c_{\alpha_{AS}}
    \end{pmatrix},
\end{equation}
where $G^0$ is the neutral Goldstone boson, and the angle $\alpha_{AS}$ is the mixing between the 2HDM pseudoscalar  and the singlet pseudoscalar $\eta_S$. In addition, the charged sector of the 2HDM+S is the same as the 2HDM, containing one pair of  charged Higgses $H^\pm$ and the Goldestone bosons $G^\pm$. 
Each of the mixing angles $\alpha,\alpha_{HS},\alpha_{hS},\alpha_{AS}$ varies in the range of 
\begin{equation}
    -\frac{\pi}{2}<\alpha_{i}<\frac{\pi}{2}.
\end{equation}
% \Shufang{Would negative angle and positive angle carry different physical consequence?}
When $\alpha_i = \pm \frac{\pi}{4}$, the mixing between the two Higgs bosons reaches maximum, and the properties of the two corresponding   scalars flip when $\frac{\pi}{4}<|\alpha_i|<\frac{\pi}{2}$. Note that the effects of different signs of the mixing angles enter only when all four mixing angles are nonzero.   When at least one mixing angle is nonzero, the properties of the Higgs bosons are independent of the sign of the mixing angles.

After the diagonalization of the Higgs mass matrices, there are  11 free parameters for the mass eigenstates:  six Higgs boson masses, $\tan\beta$ and four mixing angles. Since only the couplings between the Higgses and the SM gauge bosons enter the oblique parameters, we focus on the following nine free parameters for our study of the oblique parameters:
\begin{equation}
\underbrace{m_h=125\ {\rm GeV}, ~m_H,~m_{A},~m_{H^\pm},~\cos(\beta-\alpha),}_{\text{2HDM parameters}}~\underbrace{m_{h_S},~m_{A_S},~\alpha_{HS},~\alpha_{hS},~\alpha_{AS}}_{\text{singlet parameters}}.
\end{equation}

Using the mixing matrices, one can obtain the   couplings of physical Higgses to the gauge bosons, which are denoted by the following effective couplings
% \begin{align}
%   g_{\phi_iXX} = c_{\phi_i XX}\,{g_{h_\mathrm{SM}XX}}.
% \end{align}
\begin{equation}
g_{h_i VV}^{\mu\nu} = c_{h_i VV} i\frac{2m_V^2}{v}g^{\mu\nu},
\end{equation}
where $h_i$ represents all possible neutral CP-even states, including $h$, $H$, and $h_S$, and $V=W,Z$. The normalized couplings $c_{h_i VV}$ are shown in Table~\ref{tab:limits}.

In addition, the gauge boson can couple to two different Higgs bosons: the $Z$ boson couples to two Higgs bosons with different CP properties and the $W$ bosons couple to neutral and charged Higgs bosons. These interactions can be parametrized as
\begin{align}
    g^\mu_{\phi_i \varphi_j V} &= c_{\phi_i \varphi_j V} \,i\frac{m_V}{v}(p^\mu_{\phi_i}-p^\mu_{\varphi_j}),\\
    g^\mu_{H^- H^+ \gamma} &= c_{H^+ H^- \gamma} \,ie (p^\mu_{H^-}-p^\mu_{H^+}),\\
    g^\mu_{H^- H^+ Z} &= c_{H^+ H^- Z} \,ie\frac{c_W^2 - s_W^2}{s_W c_W} (p^\mu_{H^-}-p^\mu_{H^+}),
\end{align}
where $\phi_i$, $\varphi_j$ correspond to different types of Higgs bosons and $\varphi$ includes neutral states $\phi$ and charged Higgs $H^\pm$ \footnote{The notation of $\phi_i$ represent all possible neutral Higgs bosons, including $h,~ H,~ h_S,~ A$, and $A_S$. The notation of $\varphi_j$ represent all possible Higgs bosons, including $h,~ H,~ h_S,~ A,~ A_S$ and $H^\pm$. The general expression of $\phi_i \varphi_j V$ couplings include the couplings of $a_i h_j Z$, $a_i H^\pm W^\mp$ and $h_i H^\pm W^\mp$, where $a_i$ represents $A$ and $A_S$, and $h_j$ represents $h$, $H$ and $h_S$.}. Furthermore, the Higgs bosons can couple to gauge bosons via the quartic interactions, which are
\begin{equation}
   {g_{\varphi_i \varphi_j VV}^{\mu\nu}} =  c_{\varphi_i \varphi_j VV}\, \frac{i2 m_V^2}{v^2} g^{\mu\nu}.
\end{equation}

\begin{table}[h]
    \centering
    \begin{tabular}{llll}
    \hline
    \multicolumn{2}{l}{Benchmark Case}  &   Fixed mixing angles&   Variable mixing angles\\
    \hline
         Case-0& (2HDM alignment limit)& $c_{\beta-\alpha}=\alpha_{HS}=\alpha_{hS}=\alpha_{AS}=0$&~~~~ ---\\
         Case-I& (2HDM limit)& $\alpha_{HS}=\alpha_{hS}=\alpha_{AS}=0$& ~~~~$c_{\beta-\alpha}$\\
         Case-II& (SSM limit)& $c_{\beta-\alpha}=\alpha_{HS}=\alpha_{AS}=0$& ~~~~$\alpha_{hS}$\\
         Case-III& & $c_{\beta-\alpha}=\alpha_{hS}=\alpha_{AS}=0$& ~~~~$\alpha_{HS}$\\
         Case-IV& & $c_{\beta-\alpha}=\alpha_{hS}=\alpha_{HS}=0$& ~~~~$\alpha_{AS}$\\
         \hline
    \end{tabular}
    \caption{Five benchmark cases for the mixing angles configuration.}
    \label{tab:limits}
\end{table}

Given the complexity of the 2HDM+S scalar sectors and the appearance of multiple mixing angles, we consider five benchmark cases to disentangle the impact of each mixing angle.  For the Case-0, we have all mixing angles set to be 0, which is the 2HDM alignment limit case. For other cases, only one mixing angle is nonzero while the others are fixed to 0, as shown in Table~\ref{tab:limits}.

\begin{table}[h]
    \centering
    \resizebox{\linewidth}{!}{
    \begin{tabular}{lcccccc}
    % & & {\tiny $c_{\beta-\alpha}=\alpha_{hS}$}& {\tiny $\alpha_{hS}=\alpha_{HS}$}& {\tiny $c_{\beta-\alpha}=\alpha_{HS}$}& {\tiny $c_{\beta-\alpha}=\alpha_{hS}$}& {\tiny $c_{\beta-\alpha}=\alpha_{hS}$}\\
    % & & {\tiny $=\alpha_{HS}=\alpha_{AS}=0$}& {\tiny $=\alpha_{AS}=0$}& {\tiny $=\alpha_{AS}=0$}& {\tiny $=\alpha_{AS}=0$}& {\tiny $=\alpha_{HS}=0$}\\
    \hline
    & Couplings& Case-0&   Case-I& Case-II&    Case-III& Case-IV\\
    \hline
     & $c_{h_i VV}= R_{i1} c_\beta + R_{i2}s_\beta$& &  &  &   \\
\hline
$c_{HVV}$& $c_{\beta-\alpha}c_{\alpha_{HS}} $&    0&    $c_{\beta-\alpha}$&   0&  0& 0   \\
$c_{hVV}$& $s_{\beta-{\alpha}}c_{\alpha_{hS}} - c_{\beta-{\alpha}}s_{\alpha_{HS}}s_{\alpha_{hS}} $&    1&    $s_{\beta-\alpha}$&   $c_{\alpha_{hS}}$&   1& 1   \\
$c_{h_SVV}$& $-s_{\beta-\alpha}s_{\alpha_{hS}} - c_{\beta-\alpha}s_{\alpha_{HS}}c_{\alpha_{hS}}$&    0&    0&   $-s_{\alpha_{hS}}$&   0& 0   \\
\hline
& $c_{a_i h_j Z}  = R^A_{i1}R_{j1} + R^A_{i2}R_{j2}$\\
    \hline
         $c_{A H Z}$& $ -c_{\alpha_{AS}}c_{\alpha_{HS}}s_{\beta-{\alpha}}$& -1& $-s_{\beta-{\alpha}}$& -1& $-c_{\alpha_{HS}}$& $-c_{\alpha_{AS}}$\\
         $c_{AhZ}$& $ c_{\alpha_{AS}}\Big(c_{\beta-{\alpha}}c_{\alpha_{hS}} + s_{\beta-{\alpha}}s_{\alpha_{HS}}s_{\alpha_{hS}} \Big)$& 0&  $c_{\beta-{\alpha}}$& 0& 0& 0\\
         $c_{Ah_S Z}$& $ -c_{\alpha_{AS}}\Big(c_{\beta-{\alpha}}s_{\alpha_{hS}} - s_{\beta-{\alpha}}s_{\alpha_{HS}}c_{\alpha_{hS}} \Big)$& 0&  0& 0& $s_{\alpha_{HS}}$& 0\\
         $c_{A_S HZ}$& $s_{\alpha_{AS}}c_{\alpha_{HS}}s_{\beta-{\alpha}}$& 0&  0& 0& 0& $s_{\alpha_{AS}}$\\
        $c_{A_S h Z}$& $ -s_{\alpha_{AS}}\Big(c_{\beta-{\alpha}}c_{\alpha_{hS}} + s_{\beta-{\alpha}}s_{\alpha_{HS}}s_{\alpha_{hS}} \Big)$& 0&  0& 0& 0&  0\\
         $c_{A_S h_S Z}$& $ s_{\alpha_{AS}}\Big(c_{\beta-{\alpha}}s_{\alpha_{hS}} - s_{\beta-{\alpha}}s_{\alpha_{HS}}c_{\alpha_{hS}} \Big)$&0&  0& 0& 0& 0\\
         \hline
        & $c_{\phi_i H^\pm W^\mp}=R^{\phi}_{i2}c_\beta - R^{\phi}_{i1}s_\beta$\\
         \hline
         $c_{H H^\pm W^\mp}$& $-i c_{\alpha_{HS}}s_{\beta-{\alpha}}$& -i& $-is_{\beta-{\alpha}}$& -i& $-ic_{\alpha_{HS}}$& -i\\
         $c_{h H^\pm W^\mp}$& $i\Big(c_{\beta-{\alpha}} c_{\alpha_{hS}} + s_{\beta-{\alpha}}s_{\alpha_{HS}}s_{\alpha_{hS}} \Big)$& 0& $ic_{\beta-{\alpha}}$& 0& 0& 0\\
         $c_{h_S H^\pm W^\mp}$& $-i\Big(c_{\beta-{\alpha}} s_{\alpha_{hS}} - s_{\beta-{\alpha}}s_{\alpha_{HS}}c_{\alpha_{hS}} \Big)$& 0& 0& 0& $-is_{\alpha_{HS}}$& 0\\
         $c_{A H^\pm W^\mp}$& $c_{\alpha_{AS}}$& 1& 1& 1& 1& $c_{\alpha_{AS}}$\\
         $c_{A_S H^\pm W^\mp}$& $-s_{\alpha_{AS}}$& 0& 0& 0& 0& $-s_{\alpha_{AS}}$\\
         \hline
         &   $c_{\phi_i \phi_j VV}=R^{\phi}_{i1}R^{\phi}_{j1}+R^{\phi}_{i2}R^{\phi}_{j2}$&    &   &  &   &   \\
         \hline
         $c_{HH VV}$&  $ c^2_{\alpha_{HS}}$&   1&   1&   1&   $c^2_{\alpha_{HS}}$&1\\
         $c_{h h VV}$&  $ c^2_{\alpha_{hS}}+s^2_{\alpha_{HS}} s^2_{\alpha_{hS}}$&   1&   1&   $c^2_{\alpha_{hS}}$&   1&   1\\
         $c_{h_S h_S VV}$&  $ c^2_{\alpha_{hS}}s^2_{\alpha_{HS}}+ s^2_{\alpha_{hS}}$&   0&   0&   $s^2_{\alpha_{hS}}$&   $s^2_{\alpha_{HS}}$& 0\\
         $c_{H h VV}$&  $-\frac{1}{2} s_{2\alpha_{HS}}s_{\alpha_{hS}}$&   0&   0&   0&   0& 0\\
         $c_{H h_S VV}$&  $-\frac{1}{2} s_{2\alpha_{HS}}c_{\alpha_{hS}}$&   0&   0&   0&   $-\frac{1}{2}s_{2\alpha_{HS}}$&  0\\
         $c_{h h_S VV}$&  $-\frac{1}{2} c^2_{\alpha_{HS}}s_{2\alpha_{hS}}$&   0&   0&   $-\frac{1}{2}s_{2\alpha_{hS}}$&   0&    0\\
         $c_{A A VV}$&  $ c^2_{\alpha_{AS}}$&   1&   1&   1&   1& $ c^2_{\alpha_{AS}}$\\
         $c_{A_S A_S VV}$&  $ s^2_{\alpha_{AS}}$&   0&   0&   0& 0&     $ s^2_{\alpha_{AS}}$\\
         $c_{A A_S VV}$&  $-\frac{1}{2} s_{2\alpha_{AS}}$&   0&   0&   0&   0&    $-\frac{1}{2} s_{2\alpha_{AS}}$\\
         \hline
         $c_{H^\pm H^\mp Z}$&    &   1&  1&  1&  1&  1\\
         $c_{H^\pm H^{\mp} \gamma}$&    &   1&  1&  1&  1&  1\\
         $c_{H^\pm H^\mp VV}$&    &   1&  1&  1&  1&  1\\
         \hline
         % \multicolumn{2}{l}{Mass splitting}&     & $\Delta m_H$, $\Delta m_A$& $h, h_S$& $H, h_S$& $A, A_S$\\w
         \multicolumn{2}{l}{Relavant mixing}&  ---   & $H,h$& $h, h_S$& $H, h_S$& $A, A_S$\\
         % \multicolumn{2}{l}{Relavant Higgs contributing to $STU$}&   $H,h,A,H^\pm$  & $A,H^\pm$& $H,A,H^\pm$& $h,A,H^\pm$& $H,h,H^\pm$\\
         \hline
         
    \end{tabular}}
    \caption{The couplings between Higgs bosons and gauge bosons in 2HDM+S.  }
    \label{tab:hvcoups}
\end{table}
\begin{itemize}
\item Case-0 with $c_{\beta-\alpha}=\alpha_{HS}=\alpha_{hS}=\alpha_{AS}=0$ is the 2HDM alignment limit, where the singlet components are decoupled and 125~GeV Higgs $h$ is the same as the SM Higgs. In this case, all the couplings of the singlet Higgs bosons $h_S$, $A_S$ to SM particles are zero, and the beyond the SM (BSM) Higgs coupling $HVV$  is zero. However, the BSM Higgs bosons can still couple to gauge bosons via $AHZ$, $H H^\pm W^\mp$, $A H^\pm W^\mp$, $HHVV$,  $AAVV$  and $H^+H^-VV$ couplings. 
\item Case-I with $\alpha_{HS}=\alpha_{hS}=\alpha_{AS}=0$ is the 2HDM limit, when the singlet components are completely decoupled.  The mixing between $H$ and $h$ is parameterized by $\alpha$, as in the usual 2HDM. 
\item Case-II with $\alpha_{hS} \neq 0$ represents the case when the 125 GeV $h$ mixes with the singlet Higgs $h_S$, thus the SM-like Higgs properties are similar to those of the singlet extended SM (SSM). But the BSM doublet components $H/A$ are the same as the alignment limit of the 2HDM.
%, and $H H^\pm W^\mp$, $A H^\pm W^\mp$, $HHVV$,  $AAVV$  and $H^+H^-VV$ couplings can contribute.  
% 
\item Case-III with $\alpha_{HS} \neq 0$ represents the case when the non-SM $H$ mixes with the singlet Higgs $h_S$, while the 125~GeV Higgs $h$ is completely SM-like.
\item Case-IV with  $\alpha_{AS} \neq 0$ represents the case when $A$ mixes with the singlet pseudoscalar $A_S$, where the CP-even sector is the same as the alignment limit of the 2HDM, plus a decoupled singlet scalar $S$.
\end{itemize}

In Table.~\ref{tab:hvcoups}, we list the couplings between the Higgses and the SM gauge bosons, which are relevant for the calculation of the oblique parameters.  The general expressions are given in the second column, as well as the couplings in the individual Case-0 $-$ Case-IV. 
Since the $STU$ parameters only depend on couplings between the Higgses and gauge bosons, the fermionic couplings of the Higgs bosons are irrelevant in this study. Therefore, the contributions to the $STU$ parameters are independent of the specific structure of the Yukawa couplings.  
In particular, when the singlet CP-odd Higgs is decoupled by $\alpha_{AS}=0$,  the 2HDM+S is similar to the N2HDM (the real singlet extension of 2HDM \cite{Muhlleitner:2016mzt}). Note that, the $A_S h Z$ and $A_S h_S Z$ couplings are always zero for these benchmark cases, since multiple non-zero mixing angles are needed to couple the CP-odd singlet Higgs $A_S$ to the CP-even Higgs $h$ and $h_S$. In addition, the quartic coupling $HhVV$ is zero for the benchmark cases as well, and is non-zero only when $\alpha_{HS}$ and $\alpha_{hS}$ are both non-zero.

% %%%%%%%%%%%%%%%%%%%%%%%%%%%%%%%%%%%%%%%%%%%%%%

%%%%%%%%%%%%%%%%%%%%%%%%%%%%%%%%%%%%%%%%%%%%%%
%%%%%%%%%%%%%% section %%%%%%%%%%%%%%%%%%%%%%%
\section{Oblique parameters} 
\label{sec:EWP}

\begin{figure}[h]
    \centering
\includegraphics[width=.9\linewidth]{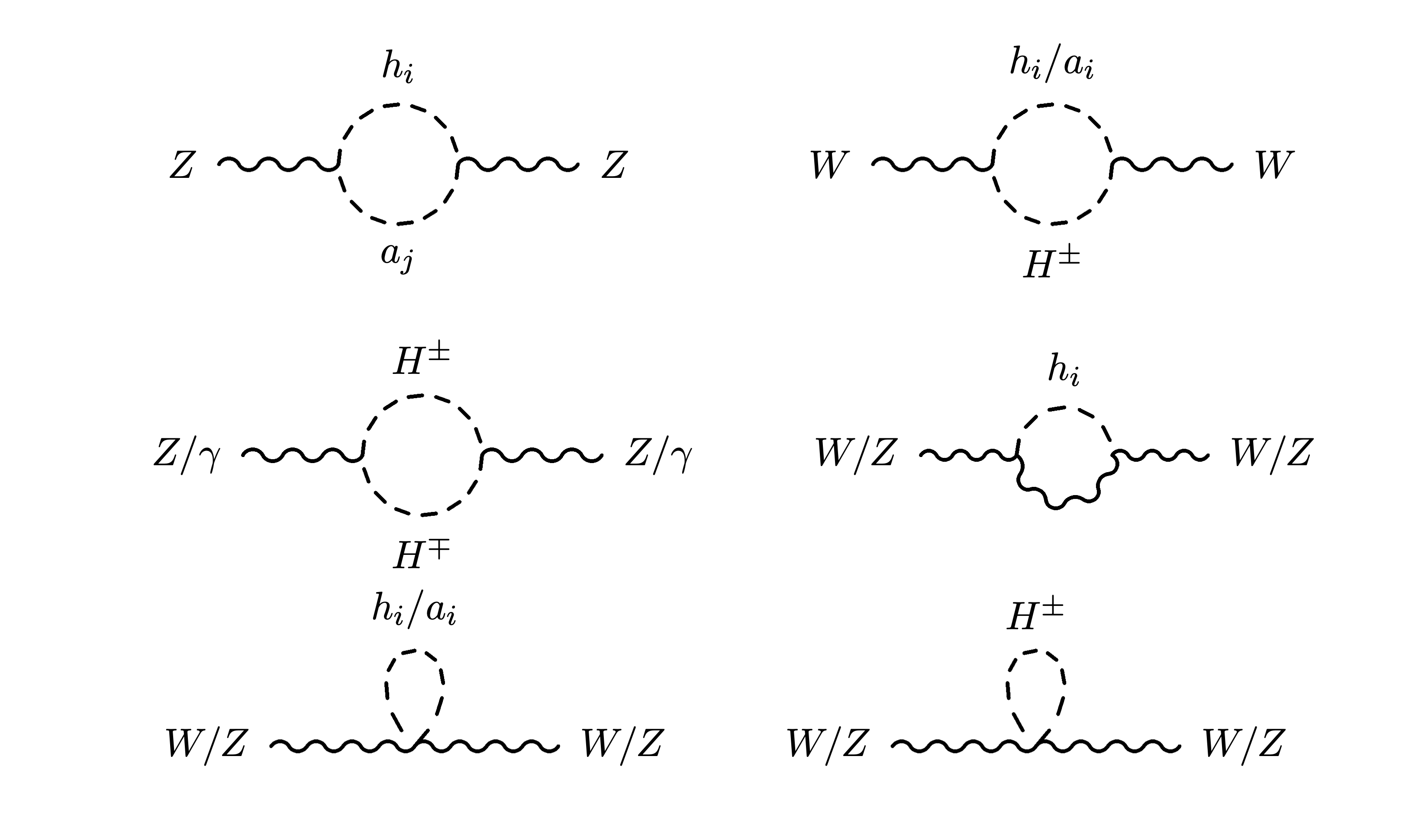}
    \caption{Feynman diagrams that contribute to the self energy of the SM gauge bosons.  }
    \label{fig:fyndig}
\end{figure}

Since the oblique parameters $STU$ are constructed with the $W$ and $Z$ self-energies~\cite{Peskin:1991sw}, as shown in Eqs.~(\ref{eq:S})-(\ref{eq:SU}), they receive contributions from the Feynman Diagrams in Fig.~\ref{fig:fyndig}. The three-point vertices (including $h_i VV$, $h_ia_j Z$, $h_i/a_i H^\pm W^\mp$ and $Z/\gamma H^\pm H^\mp$), as well as the four-point vertices (including $h_i h_i VV$, $a_i a_i VV$ and $H^\pm H^\mp VV$), contribute to the gauge bosons self-energies.

The contributions to the $STU$ parameters from various Higgses can be found in Ref.~\cite{Grimus:2008nb}. Using those expressions, the $STU$ parameters in 2HDM+S are given by
    \begin{eqnarray}
         S&=&\frac{1}{24\pi}\Bigg[  (2s_W^2-1)^2G(m_{H^\pm}^2,m_{H^\pm}^2,m_Z^2)
        +\sum_{i,j} |{c_{a_i h_j Z}}|^2 G(m_{a_i}^2,m_{h_j}^2,m_Z^2) \nonumber \\
        &+&\sum_{i=1}^3 {c_{h_ih_iVV}} \ln(m_{h_i}^2) +\sum_{i=1}^2 {c_{a_i a_iVV}} \ln(m_{a_i}^2)-2\ln(m_{H^\pm}^2) -\ln(m_{h_\text{ref}}^2) \nonumber \\
        &+&\sum_{i=1}^3 |c_{h_i VV}|^2  \hat G(m_{h_i}^2,m_Z^2)-\hat G(m_{h_\text{ref}}^2,m_Z^2)            \Bigg],
     \label{eq:gen_S} 
    \end{eqnarray}
     \begin{eqnarray}
        T&=&\frac{1}{16\pi s_W^2 m_W^2}\Bigg[ \sum_{i=1}^3 {|c_{h_i H^\pm W^\mp}|^2} F(m_{H^\pm}^2,m_{h_i}^2) + \sum_{i=1}^2 {|c_{a_i H^\pm W^\mp}|^2}F(m_{H^\pm}^2,m_{a_i}^2)
        \nonumber \\
          &-&  \sum_{i,j} {|c_{a_i h_j Z}|^2} F(m_{a_i}^2,m_{h_j}^2) +3 \sum_{i=1}^3 |c_{h_i VV}|^2 \left(F(m_Z^2,m_{h_i}^2)-F(m_W^2,m_{h_i}^2) \right) \nonumber \\
          &-& 3 \left(F(m_Z^2,m_{h_\text{ref}}^2)-F(m_W^2,m_{h_\text{ref}}^2) \right)
        \Bigg], 
    \label{eq:gen_T}
    \end{eqnarray}
   \begin{eqnarray}    
        U&=&\frac{1}{24\pi}\Bigg[  \sum_{i=1}^3 {|c_{h_i H^\pm W^\mp}|^2}G(m_{H^\pm}^2,m_{h_i}^2,m_W^2)+ \sum_{i=1}^2 {|c_{a_i H^\pm W^\mp}|^2}G(m_{H^\pm}^2,m_{a_i}^2,m_W^2) \nonumber \\
        &-&(2s_W^2-1)^2 G(m_{H^\pm}^2,m_{H^\pm}^2,m_Z^2)-  \sum_{i,j} {|c_{a_i h_j Z}|}^2 G(m_{a_i}^2,m_{h_j}^2,m_Z^2) \nonumber \\
        &+&\sum_{i=1}^3 |c_{h_i VV}|^2\left(\hat G(m_{h_i}^2,m_W^2)-\hat G(m_{h_i}^2,m_Z^2) \right)-\hat G(m_{h_\text{ref}}^2,m_W^2)+\hat G(m_{h_\text{ref}}^2,m_Z^2) \Bigg], 
  \label{eq:gen_U}
\end{eqnarray}
where the $m_{h_\text{ref}}=125$~GeV is the reference mass of the SM Higgs.  Functions $F$, $G$ and $\hat{G}$ can be found in Eqs.~\eqref{eq:fij},~\eqref{eq:Gijq} and \eqref{eq:Ghat} in Appendix~\ref{sec:append}.% \Shufang{Add the reference or Eq number if in appendix.}

For the $T$ parameter, the contributions from the quartic couplings cancel, since $\varphi_i\varphi_i W W$ are the same as $\varphi_i\varphi_i ZZ$ and $T$ observable is defined by the self-energy difference between $W$ boson and $Z$ boson (see Eq.~\eqref{eq:T}). Thus, the $T$ observable only receives the contribution from the $h_i VV$, $a_i h_j Z$ and $a_i/h_i H^\pm W^\mp$ couplings. Furthermore, the $S$ parameter mainly represents the $Z$ boson self-energy, and receives contributions from the $ZH^\pm H^\mp$ interaction via $G(m_{H^\pm}^2,m_{H^\pm}^2,m_Z^2)$ and the $a_i h_j Z$ interaction via $G(m_{a_i}^2,m_{h_j}^2,m_Z^2)$. In addition, the quartic couplings $h_i h_i VV$, $a_i a_i VV$ and $H^\pm H^\pm VV $ enter into the $S$ parameter via the logarithmic functions.
For the $U$ parameter, the contributions of the quartic interactions cancel again. Furthermore, the $U$ parameter is related to the dim-8 operator, which is usually suppressed. Therefore,  in our discussion below, we mostly focus on the $S$ and $T$ parameters, which are more sensitive to the BSM effects.

The experimental measurements for the electroweak precision observables yield the following best-fit values of $STU$ \cite{stupdg:2024} for $m_{h_\text{ref}}=125$ GeV: %\Shufang{125 GeV?}%\cite{Haller:2018nnx}:
\begin{equation}
    \begin{matrix}
        S^\mathrm{exp}=-0.04, & T^\mathrm{exp}=0.01,& U^\mathrm{exp}=-0.01,\\
        \Delta S=0.10, & \Delta T=0.12,& \Delta U=0.09,\\
        \text{corr}(S,T)=+0.93, & \text{corr}(S,U)=-0.70,& \text{corr}(T,U)=-0.87, \label{eq:STU_res}
    \end{matrix}
\end{equation}
% \Wei{\url{https://pdg.lbl.gov/2024/reviews/rpp2024-rev-standard-model.pdf}, here are lastest STU values}
where $\text{corr}(S,T)$,  $\text{corr}(S,U)$ and $\text{corr}(T,U)$ are the correlation coefficients between the $S$, $T$ and $U$. The contributions to the oblique parameters $STU$  in 2HDM+S, i.e.,  Eqs.~\eqref{eq:gen_T},~\eqref{eq:gen_S} and \eqref{eq:gen_U},  can be used to construct the $\chi^2$ value~\cite{Baak:2011ze,Haller:2018nnx}, 
\begin{equation}
    \chi^2_{STU}=\begin{pmatrix} S-S^\mathrm{exp}, & T-T^\mathrm{exp}, & U-U^\mathrm{exp} \end{pmatrix}   \cdot  \textbf{cov}^{-1} \cdot \begin{pmatrix} S-S^\mathrm{exp} \\ T-T^\mathrm{exp} \\ U-U^\mathrm{exp} \end{pmatrix},
\end{equation}
where
\begin{equation}
    \textbf{cov}=\begin{pmatrix}
        \Delta {S}^2 & \text{corr}(S,T) \Delta S\Delta T &\text{corr}(S,U) \Delta S\Delta U\\
        \text{corr}(S,T) \Delta S\Delta T  & \Delta {T}^2 &\text{corr}(T,U) \Delta T\Delta U\\
        \text{corr}(S,U) \Delta S\Delta U  & \text{corr}(T,U)\Delta {T}\Delta {U}  &\Delta U^2
    \end{pmatrix}.
\end{equation}

The two-dimensional fit to the $STU$ parameters at 95\% C.L. corresponds to $\Delta\chi^2 = \chi^2_{STU} - \chi^2_{STU}|_{\text{minimal}} < 5.99$.
% %%%%%%%%%%%%%%%%%%%%%%%%%%%%%%%%%%%%%%%%%%%%%%

%%%%%%%%%%%%%%%%%%%%%%%%%%%%%%%%%%%%%%%%%%%%%%
%%%%%%%%%%%%%% section %%%%%%%%%%%%%%%%%%%%%%%
\section{Five Benchmark Cases}
\label{sec:res}

% \subsection{The impact of the singlet admixture on $STU$ constraint}
In the 2HDM, the $STU$ parameters play an important role in constraining the mass splittings between the BSM neutral Higgses and the charged Higgses $H^\pm$. In 2HDM+S, the singlet field enters via the mixing, which further changes the dependence of the $STU$ parameters on model parameters. In this section, we explore the impacts of electroweak constraints on the mixing angles, $\beta-\alpha$, $\alpha_{hS}$, $\alpha_{HS}$, and $\alpha_{AS}$, as well as various mass splittings.  For convenience, we define the following mass splittings, which are relevant for the $STU$ constraints,
\begin{eqnarray}
    \Delta m_{H}& = m_H - m_{H^\pm}, \  \ \ 
    \Delta m_A = m_A - m_{H^\pm},\nonumber\\
    % \Delta m_{hS} &= m_{h_S} - m_{h_{125}}\\
    \Delta m_{h_S} &= m_{h_S} - m_{H^\pm}, \ \ \ 
    \Delta m_{A_S} = m_{A_S} - m_{H^\pm}.
\end{eqnarray}
% \Shufang{check the notation of $\Delta m _{h_S}$ }.

\subsection{Case-0}
As a starting point, we study the simplest Case-0 (the 2HDM alignment limit)  with $c_{\beta-\alpha}=\alpha_{HS}=\alpha_{hS}=\alpha_{AS}=0$. According to Table~\ref{tab:hvcoups}, the non-zero couplings in this case are
\begin{equation}
    c_{hVV},~c_{AHZ},~ c_{HH^\pm W^\mp},~ c_{A H^\pm W^\mp},~c_{ZH^\pm H^\mp},~c_{hhVV},~c_{HHVV},~c_{AAVV},~c_{H^\pm H^\pm VV},
    \label{eq:case0coupling}
\end{equation}
with norm 1. The 125~GeV Higgs $h$ is the SM Higgs 
and singlet Higgs bosons $h_S$ and $A_S$ both decouple.  
The doublet BSM Higgses $H$, $A$ and $H^\pm$ enter via $AHZ$, $H H^\pm W^\mp$ and $A H^\pm W^\mp$ interactions and mainly contribute to the terms involving $F$ functions in the $T$ parameter. In addition,  $Z H^\pm H^\mp$ and  quartic interactions $HHVV$, $AAVV$ and $H^\pm H^\mp VV$ contribute to the $S$ parameter.  Consequently, the masses of $H$, $A$ and $H^\pm$ are relevant for the oblique parameters, while the singlet Higgs masses $m_{h_S}$ and $m_{A_S}$ are irrelevant. 
%This case is the same as the alignment limit of the 2HDM  with $c_{\beta-\alpha}=0$. 

\begin{figure}[h]
    \centering
    \includegraphics[width=.5\linewidth]{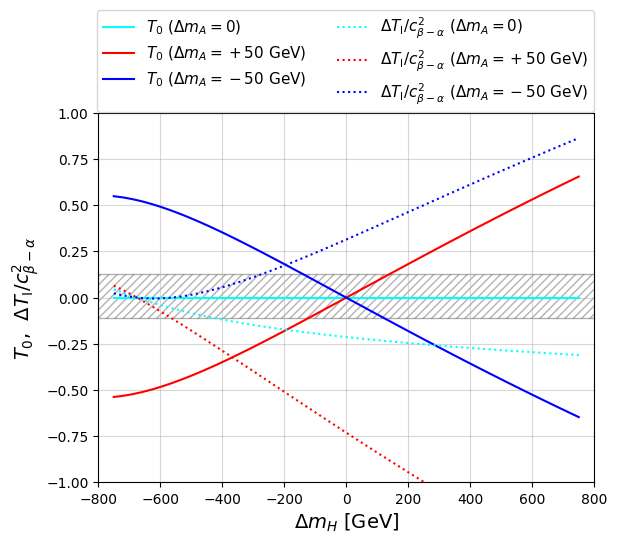}\includegraphics[width=.5\linewidth]{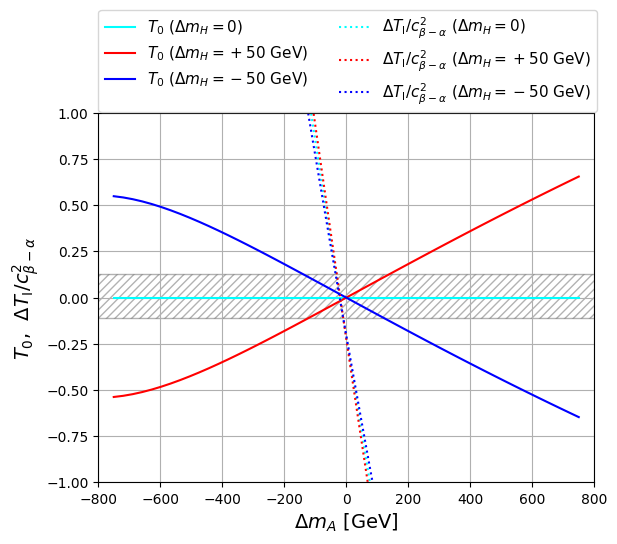}
    \caption{$T_0$ (solid lines) and $\Delta T_\text{I}/c^2_{\beta-\alpha}$ (dotted lines) with varying $\Delta m_H$ (left) and $\Delta m_A$ (right). Cyan, red, and blue lines are for $\Delta m_{A,H}=$0, + 50~GeV and $-$50~GeV, respectively. The grey hatch area is the 1$\sigma$ region of the $T$ observable. $m_{H^\pm}$ is chosen to be 800 GeV.
 }
    \label{fig:t0t1}
\end{figure}

% \begin{figure}
% \centering
% \includegraphics[width=.5\linewidth]{figs/tmh1tev.png}
% \end{figure}

The $T$ and $S$ parameters in this case, denoted as $T_0$ and $S_0$, are given by~\cite{Baak:2011ze}
\begin{align}
     T_0 &= \frac{1}{16\pi s^2_W m_W^2}[F(m_{H^\pm}^2, m_H^2) - F(m_{A}^2, m_H^2)   + F(m_{H^\pm}^2, m_A^2)],\label{eq:t0}\\
     S_0 &= \frac{1}{24\pi}[(2s_W^2-1)^2G(m_{H^\pm}^2,m_{H^\pm}^2,m_Z^2) + G(m_A^2, m_H^2, m_Z^2) +  \ln \left( \frac{m_H^2}{m_{H^\pm}^2}  \right) + \ln \left( \frac{m_A^2}{m_{H^\pm}^2}  \right) ],%+ \text{const.}
     \label{eq:s0}
\end{align}
The values of $T_0$ with varying $\Delta m_H$ and $\Delta m_A$ are presented by the solid lines in Fig.~\ref{fig:t0t1}. The left panel is for varying $\Delta m_H$ with fixed $\Delta m_A=0,\ \pm50$~GeV, and the right panel is for varying $\Delta m_A$ with fixed $\Delta m_H=0,\ \pm50$~GeV. As indicated by Eq.~(\ref{eq:t0}), $T_0$ is exactly zero when $\Delta m_H=0$ or $\Delta m_A=0$. The grey hatch area is the  1$\sigma$ region of the electroweak precision observable fit to the $T$ parameter. $T_0$ is also symmetric under the exchange of $m_H$ and $m_A$.  Therefore, the $T_0$ dependence on $\Delta m_H$ in the left panel is the same as the 
$T_0$ dependence on $\Delta m_A$ in the right panel.
$T_0$  increases as $\Delta m_H$ ($\Delta m_A$) increases for $\Delta m_A>0$  ($\Delta m_H >0$), while decreases for the opposite sign of $\Delta m_A$ ($\Delta m_H$).  Furthermore, $T_0>0$ when both $\Delta m_H$ and $\Delta m_A$ have the same sign, and $T_0<0$ when $\Delta m_H$ and $\Delta m_A$ have the opposite signs.

The $S_0$ parameter, however, is not zero even when both $\Delta m_A$ and $ \Delta m_H$ are zero. The contributions from $G(m_i^2, m_j^2, m_k^2)$, however, are typically very small. The main contributions to $S_0$ come from the logarithmic terms $\ln(m_{H,A}^2/m_{H^\pm}^2)$. For $\Delta m_{H,A}$ in the range of $\pm 700 $~GeV, $|S_0|<0.15$ is within 1 $\sigma$ range of the fitted value.

\begin{figure}[h]
\centering
% \subfigure[$\Delta m_H$ vs $\Delta m_A$]
\includegraphics[width=.7\linewidth]{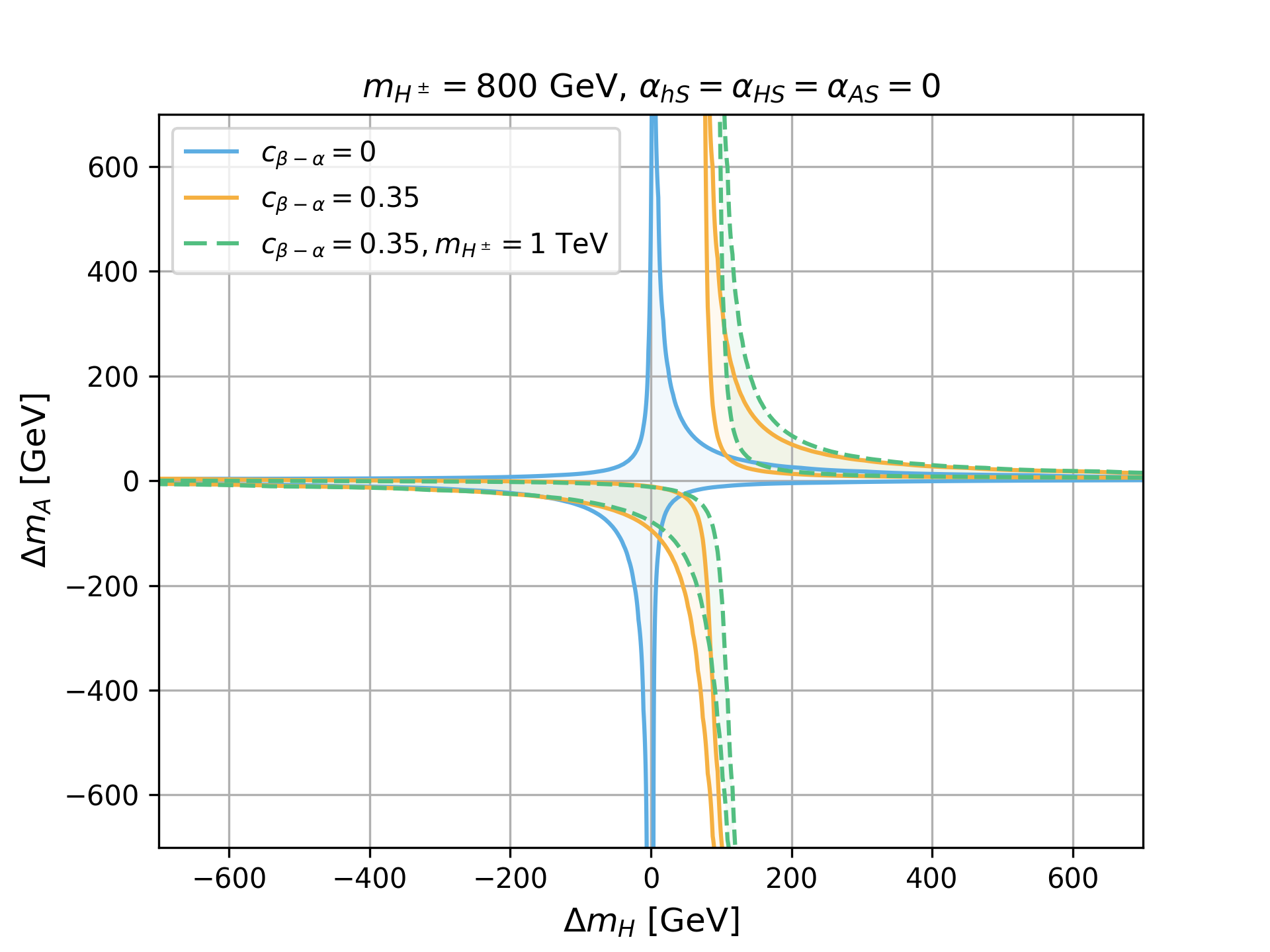}
\caption{The 95\% C.L. allowed region from $STU$ constraints in the plane $\Delta m_H$ vs. $\Delta m_A$ with $c_{\beta-\alpha}=0$ (solid blue region) and $0.35$ (solid orange region) for $m_{H^\pm} = 800$~GeV. The other parameters are $\alpha_{HS}=\alpha_{hS}=\alpha_{AS}=0$. For $m_{H^\pm} = 1000$~GeV, the allowed region for $c_{\beta-\alpha}=0$ is approximately the same to $m_{H^\pm} = 800$~GeV,  while the region for $c_{\beta-\alpha}=0.35$ are shown by the   green regions.}
\label{fig:mamh}
\end{figure}

Fig.~\ref{fig:mamh} shows the 95\% C.L. allowed region from the $STU$ constrains in the $\Delta m_H$ vs. $\Delta m_A$ plane. The blue region corresponds to the {Case-0} with $c_{\beta-\alpha}=\alpha_{HS}=\alpha_{hS}=\alpha_{AS}=0$ and $m_{H^\pm} = 800$~GeV, which centers around $\Delta m_H=0$ or $\Delta m_A=0$.
% \Wei{stop here} 
Due to the positive correlation between $S$ and $T$ observables, the area with positive $T$ is preferred. Therefore, the allowed regions with the same signs of $\Delta m_A$ and $\Delta m_H$ are bigger comparing to the allowed regions with the opposite signs.  

\begin{figure}[h]
    \centering
    % \hfill
    % \subfigure[]{\label{fig:cbamh}
    \includegraphics[width=0.5\linewidth]{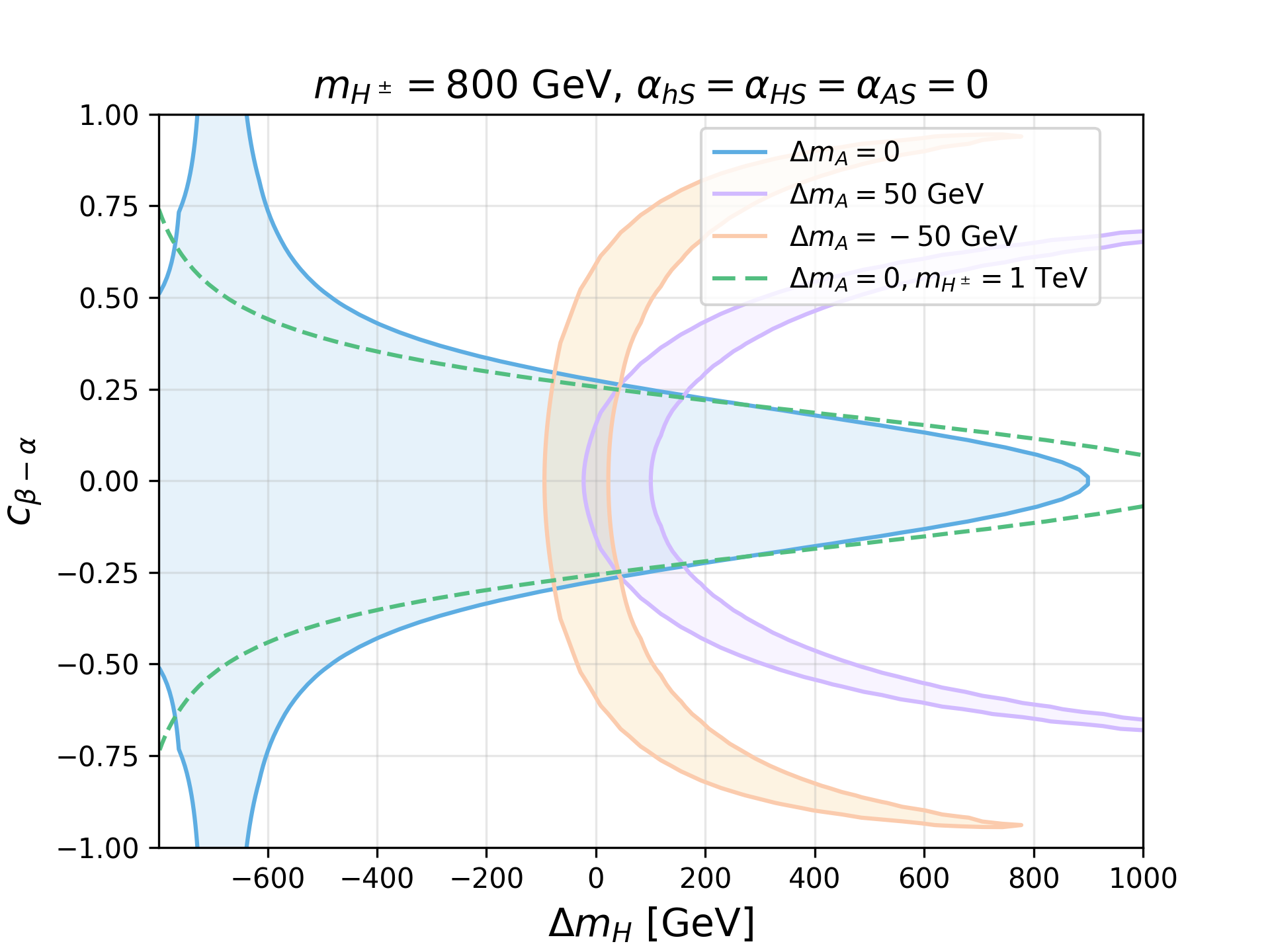}%}%\hfill
        % \subfigure[]{\label{fig:cbama}
        \includegraphics[width=0.5\linewidth]{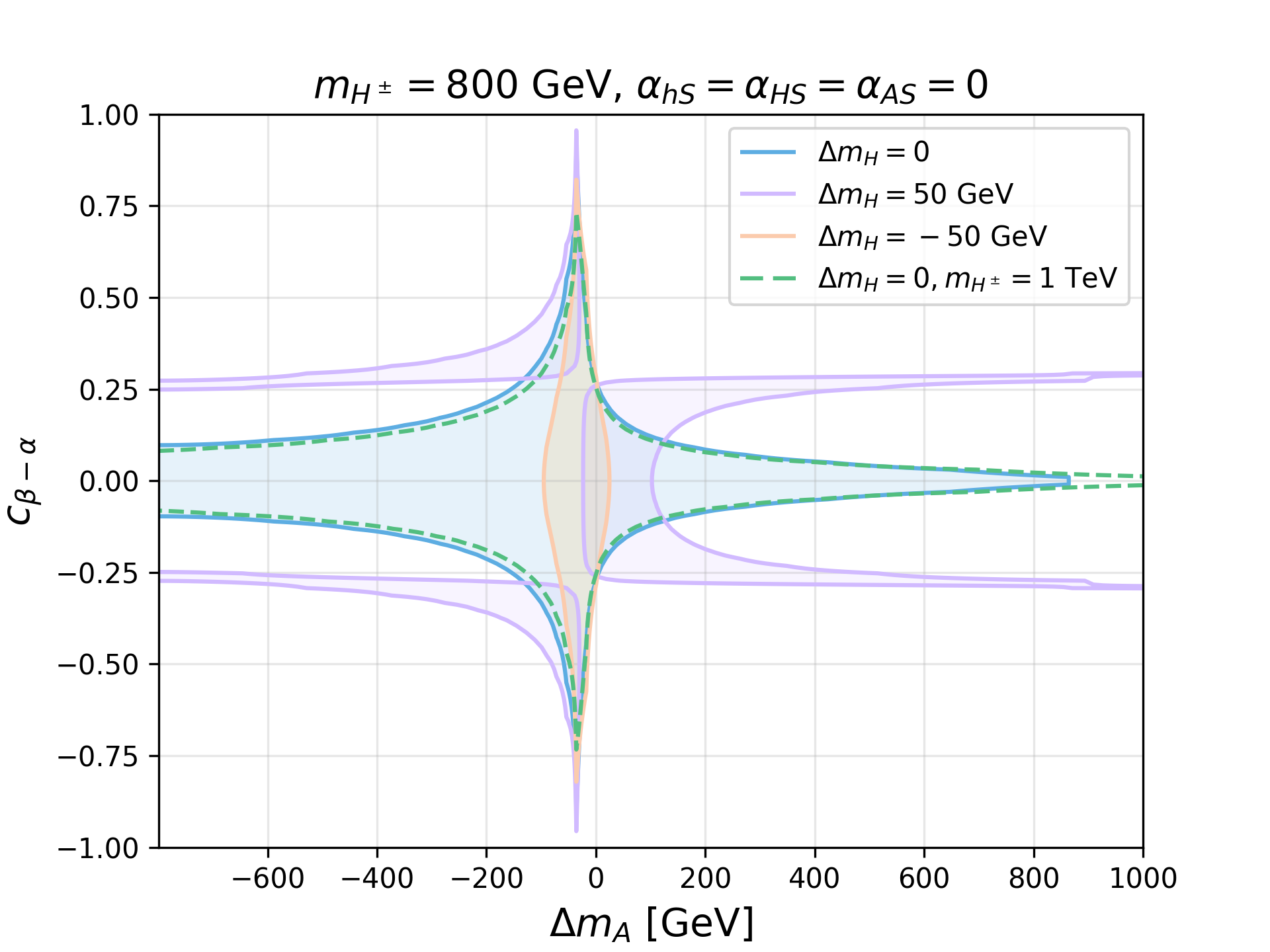}%}%$c_{\beta-\alpha}$ vs $\Delta m_H$ with $\Delta m_A=0$ and $\pm 50$~GeV$c_{\beta-\alpha}$ vs $\Delta m_A$ with $\Delta m_H=0$ and $\pm 50$~GeV
    \caption{The 95\% C.L. allowed region via $STU$ constraints on $c_{\beta-\alpha}$ vs. $\Delta m_H$ (left) and $\Delta m_A$ vs $c_{\beta-\alpha}$ (right). The other parameters are chosen as $\alpha_{HS}=\alpha_{hS}=\alpha_{AS}=0$ and $m_H^\pm = 800$~GeV (solid curves). Blue, purple and orange regions correspond to $\Delta m_{A,H}=0,\ 50,\ -50$ GeV, respectively.  Green dashed curves are for $m_{H^\pm} = 1000$~GeV and $\Delta m_{A,H=0}$.}% \Shufang{Delta mA in the right plot in x-axis.} \Wei{remove the (a)(b), just use left and right pannel during description easily.}}
    \label{fig:2hdmcase}
\end{figure}

In Figs.~\ref{fig:2hdmcase}, the 95\% C.L. allowed regions under the $STU$ constraints are shown in the $\Delta m_{A,H}$ vs. $c_{\beta-\alpha}$  plane. 
For $c_{\beta-\alpha}=0$, the 95\% C.L. fit to the $STU$ parameters gives $\Delta m_{A,H} \lesssim 900$~GeV with $\Delta m_{H,A} = 0$ for $m_{H^\pm}=800$ GeV (blue region).The upper limits on $\Delta m_{A,H}$ come from the logarithm contributions.
%CL{($\ln\left(\frac{m_{H,A}^2}{m_{H^\pm}^2} \right) = 2\ln(1+\frac{\Delta m_{H,A}}{m_{H^\pm}})$)} to the $S$ parameter.\Wei{ can we have a rough estimation of the upper limit value based on the formulae?}
% \Wei{$c_{\beta-\alpha}$ gets the larger allowed region around $\Delta m_H = -770$ GeV(800?) because of the cancellation xxx... It looks the explanation is in case-I. we can have a few here firstly. }
%\Wei{Here is only for cba=0}
These upper limits of $\Delta m_{A,H}$ vary with the benchmark value of $m_{H^\pm}$ and increase as ${H^\pm}$ becomes heavier, as indicated by the green dashed curve for $m_{H^\pm}=1000$ GeV. %\Wei{why? here we only show the results without any explanation}\CL{$\frac{\Delta m_{H,A}}{m_{H^\pm}}$ is suppressed}\Wei{from S0}. 
For non-zero values of $\Delta m_{A,H}=\pm 50$ GeV, the allowed range of $\Delta m_{H,A}$ is much smaller, as shown by the regions with the purple and orange boundary curves in  Figs.~\ref{fig:2hdmcase}.

\subsection{Case-I: $c_{\beta-\alpha}\neq0$
%($c_{\beta-\alpha}\neq0$ and $\alpha_{HS}=\alpha_{hS}=\alpha_{AS}=0$)
}

In the {Case-I} ($c_{\beta-\alpha}\neq0$ and $\alpha_{HS}=\alpha_{hS}=\alpha_{AS}=0$), the singlet fields decouple completely, and the model is the same as the 2HDM. In particular, we have the following non-zero couplings in addition to those shown in Eq.~(\ref{eq:case0coupling})
\begin{equation}
    %c_{hVV},~c_{HVV},~c_{AHZ},~c_{AhZ},~ c_{A H^\pm W^\mp}, c_{HH^\pm W^\mp},~c_{hH^\pm W^\mp},
c_{HVV},~c_{AhZ}, ~c_{hH^\pm W^\mp},
\end{equation}
which are proportional to $c_{\beta-\alpha}$ and provide additional contributions to the $STU$ parameters. Similar to the Case-0, the singlet masses $m_{h_S}$ and $m_{A_S}$ are irrelevant, and only the doublet-like Higgs masses $m_H$, $m_A$ and $m_{H^\pm}$ enter.  
The $STU$ constraints of 2HDM have been studied in the literature~\cite{Haller:2018nnx,Gorbahn:2015gxa}. 
The $T$ observable in the Case-I is
\begin{eqnarray}
        T_\text{I} &=& T_0 + \Delta T_\text{I} \\
        \Delta T_\text{I}&=& \frac{c^2_{\beta-\alpha}}{ 16\pi s^2_W m_W^2 } \{ F(m_h^2, m_{H^\pm}^2) - F (m_h^2, m_A^2) - [F(m_H^2, m_{H^\pm}^2) - F (m_H^2, m_A^2)]  \nonumber \\
        &&- 3  [F(m_h^2, m_{Z}^2) - F (m_h^2, m_{W^\pm}^2)] + 3[F(m_H^2, m_{Z}^2) - F (m_H^2, m_{W^\pm}^2)] \}  
.\label{eq:t11}
\end{eqnarray}
Comparing to the Case-0, the additional contribution of $\Delta T_\text{I}$  is proportional to $c_{\beta-\alpha}$, which is non-zero even for  $\Delta m_A = 0$.

In Fig.~\ref{fig:t0t1}, we show the value of $T_0$ (solid curves) and $\Delta T_\text{I}/c_{\beta-\alpha}^2$ (dashed curves) for different values of $\Delta m_H$ and $\Delta m_A$.  The left panel shows that for $\Delta m_H=-675$ GeV, which corresponds to $m_H= m_h=125$ GeV, $\Delta T_\text{I} = 0$.   Right panel shows that $\Delta T_\text{I}$ has the opposite (same) sign of $T_0$ for positive (negative) $\Delta m_H$, except for a small negative $\Delta m_A$ region.

The 95\% C.L.  $STU$ allowed parameter space in $\Delta m_H$ vs $\Delta m_A$ plane are shown in Fig.~\ref{fig:mamh} for $c_{\beta-\alpha}=0.35$ (orange).  The allowed regions shift to the right ($\Delta m_H > 0$), given the cancellation between $T_0$ and $\Delta T_\text{I}$. In particular, the $\Delta m_A = 0$ point with $\Delta m_H\sim$100~GeV and $c_{\beta-\alpha}=0.35$ would be excluded, since $T_0$ is zero and cannot eliminate the non-zero $\Delta T_\text{I}$. The regions enclosed by the green dashed curves are for $m_{H^\pm}=1000$ GeV, which is close to the orange regions of $m_{H^\pm}=800$ GeV.

Left panel of Fig.~\ref{fig:2hdmcase} shows the 95\% C.L. $STU$ allowed parameter space in  $c_{\beta-\alpha}$ vs $\Delta m_H$ for various $\Delta m_A$. The allowed regions are symmetric with respect to $c_{\beta-\alpha}=0$ given the $c_{\beta-\alpha}^2$ dependence. For $\Delta m_A=0$ (region enclosed by the solid blue curve), all values of $c_{\beta-\alpha}$ is allowed at $m_{H}=125$~ GeV:  $T_0=0$ since $\Delta m_A = 0$, and $\Delta T_\text{I}=0$ for $m_H = m_{h}=125$~GeV.  The allowed regions shrink for larger $|m_H-m_{h_{125}}|$.
The green dashed line indicates the impact of the value of $m_H^\pm$.  For non-zero $\Delta m_A$,  the non-zero $T_0$ could be canceled by $\Delta T_\text{I}$.  The allowed regions favor mostly positive $\Delta m_H$, as shown by the regions enclosed by the purple curves and orange curves.    Since the absolute value of $\Delta T_\text{I}$ is larger when $\Delta m_A$ is positive as shown in  Fig.~\ref{fig:t0t1}, the allowed regions with positive $\Delta m_A$ favor smaller $|c_{\beta-\alpha}|$.  The 2HDM non-alignment case has been studied in \cite{Gorbahn:2015gxa}, which did not cover the case with much larger mass splittings.

The right panel of Fig.~\ref{fig:2hdmcase} shows the 95\% C.L. $STU$ allowed parameter space in  $c_{\beta-\alpha}$ vs $\Delta m_A$ for various $\Delta m_H$.  For $\Delta m_H=0$ (region enclosed by the solid blue curves), relatively large region of $c_{\beta-\alpha}$ is allowed for $\Delta m_A \sim -30$ GeV, when $\Delta T_\text{I} \sim 0$ and $S_0$ and $T_0$ are small.  The allowed regions of $c_{\beta-\alpha}$ are smaller for $\Delta m_A >0$ comparing to $\Delta m_A <0$, since $|\Delta T_\text{I}|$ is larger for positive $\Delta m_A$.  Also note that for negative $\Delta m_H=-50$ GeV, only a narrow range of $\Delta m_A$ around $-30$ GeV is allowed.  This is because $\Delta T_\text{I}$ has the same signs as $T_0$ for negative $\Delta m_H$.  Therefore, only small values of $\Delta m_A$ are allowed. 
However, for positive $\Delta m_H=50$ GeV, $\Delta T_\text{I}$ and $T_0$ have opposite signs. A wide range of $\Delta m_A$ is allowed: $|c_{\beta-\alpha}|\lesssim 0.25$ for $\Delta m_A>0$, and $|c_{\beta-\alpha}| \gtrsim 0.25$ for $\Delta m_A<0$.

% \newpage
\subsection{Case-II: $\alpha_{hS}\neq0$}

\begin{figure}[h]
    \centering
    \includegraphics[width=.5\linewidth]{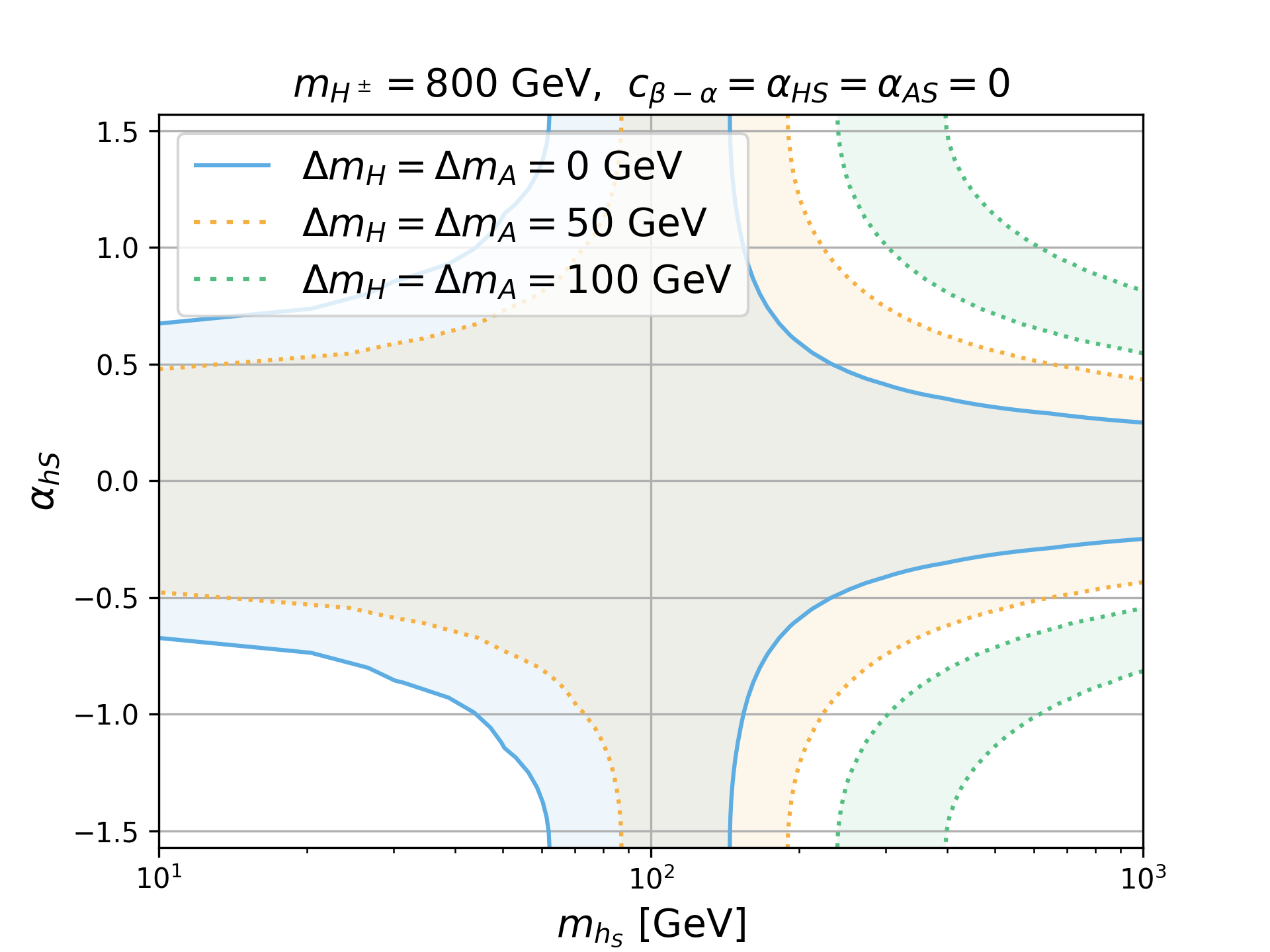}\includegraphics[width=.5\linewidth]{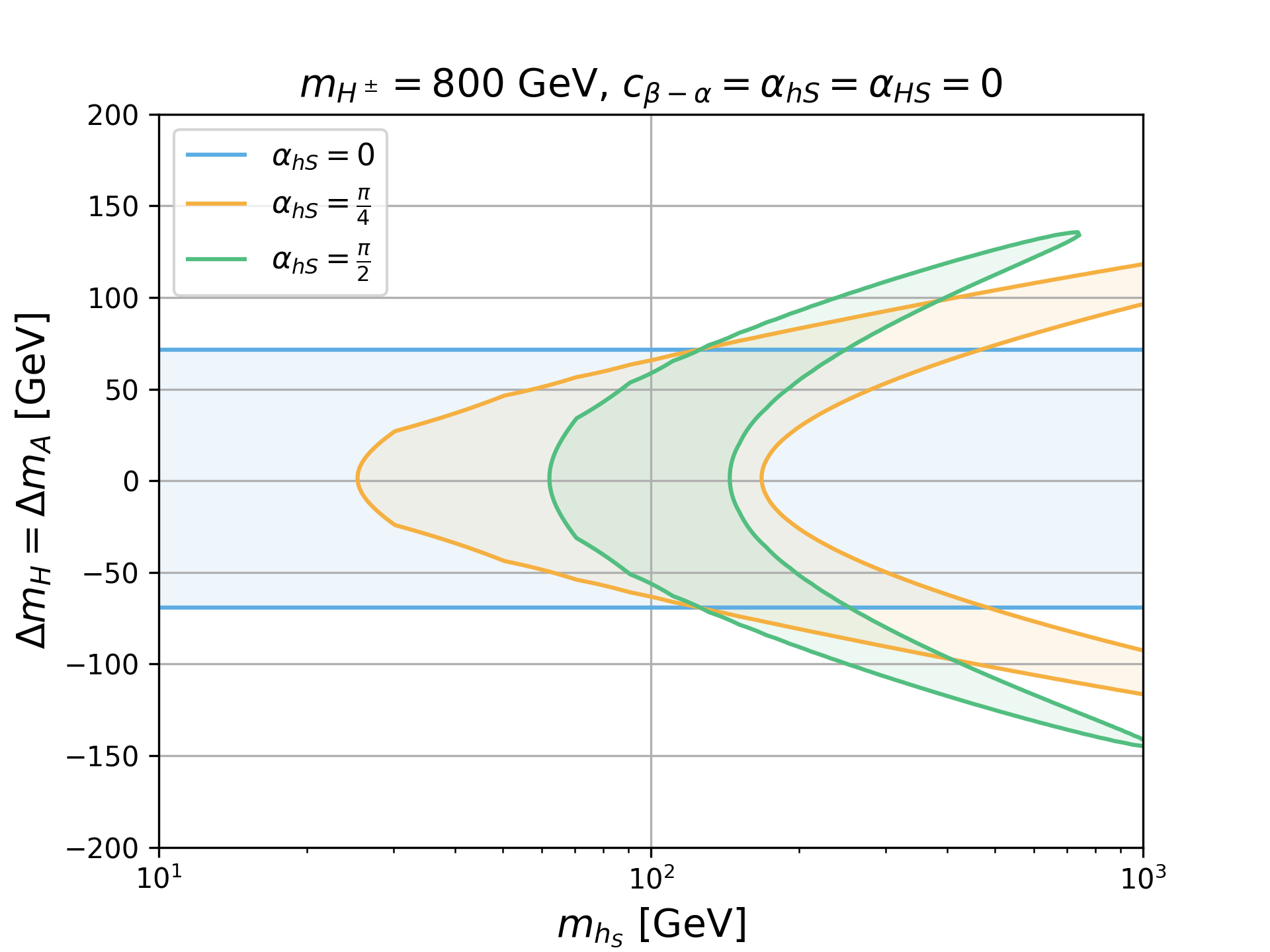}
    \caption{95\% C.L. $STU$ constraints on the parameter space of $m_{h_S}$, $\alpha_{hS}$ and $\Delta m_A$. The left panel is for $m_{h_S}$ vs $\alpha_{hS}$ with varying $\Delta m_A$. The blue, orange and green regions are for  $\Delta m_A = \Delta m_H=0$, 50,~100~GeV, respectively. The right panel is for $m_{h_S}$ vs $\Delta m_{H,A}$ with varying $\alpha_{hS}$. The blue, orange and green regions are for  $\alpha_{hS}=0$, $\frac{\pi}{4}$ and $\frac{\pi}{2}$, respectively. For both panel,  we set $m_{H^\pm}= 800$ GeV and  $c_{\beta-\alpha}=\alpha_{HS}=\alpha_{AS}=0$. }
    \label{fig:mhsahs}
\end{figure}
{In the Case-II (e.g. $\alpha_{hS}\neq0$ and $c_{\beta-\alpha}=\alpha_{HS}=\alpha_{AS}=0$), the 125~GeV Higgs $h$ mixes with the singlet-like Higgs $h_S$, and the $h_S VV$ coupling is proportional to $s_{\alpha_{hS}}$, which is the only non-zero trilinear coupling between Higgs and  gauge bosons,   in addition to those in Eq.~(\ref{eq:case0coupling}). 
The $A h_S Z$ and $h_S H^\pm W^\mp$ couplings are still zero, which means $\alpha_{hS}$ cannot connect the $h_S$ with the BSM doublet-like Higgses $H$, $A$ or $H^\pm$. This case is similar to the singlet extension of the SM (SSM), where the singlet Higgs only mixes with the SM Higgs $h$.} Therefore, the $STU$ parameters receive additional contribution via loops with $h_SVV$ vertices, with the singlet Higgs mass $m_{h_S}$ entering. %For the 2HDM, the additional BSM Higgs coupling $c_{HVV}=\cos(\beta-\alpha)$ would be suppressed by the alignment limit $\cos(\beta-\alpha)=0$, and the $HVV$ contribution on the $STU$ observables would be tiny. In contrast, the 2HDM+S still has the $h_SVV$ contribution when the $h_SVV$ coupling is dominated by the mixing angle $\alpha_{hS}$. For instance, 
The $S$ and $T$ parameters are given by
\begin{align}
& S_\text{II}= S_0+\Delta S_\text{II},\ \ \ T_\text{II}=T_0 + \Delta T_\text{II},\\
& \Delta S_\text{II}=\frac{1}{24\pi}s^2_{\alpha_{hS}} \Big[  \ln \left( \frac{m_{h_S}^2}{m_{h_{125}}^2}\right) + \hat{G}(m_{h_S}^2, m_Z^2) - \hat{G}(m_{h_{125}}^2, m_Z^2) \Big],\label{eq:s3}\\
& \Delta T_\text{II}=\frac{1}{16\pi s_W^2 m_W^2}3s^2_{\alpha_{hS}}\Big[F(m_Z^2,m_{h_S}^2)-F(m_W^2,m_{h_S}^2) - F(m_Z^2,m_{h_{125}}^2)+F(m_W^2,m_{h_{125}}^2) \Big]   \label{eq:t3}
\end{align}
The expression for the function $\hat{G}$ can be found in Eq.~(\ref{eq:Ghat}).
Note that $\Delta T_\text{II}$ and $\Delta S_\text{II}$ are proportional to $s^2_{\alpha_{hS}}$, and both terms vanish when $m_{h_S}=m_{h_{125}}$. $\Delta S_\text{II}$ is in general suppressed, while $\Delta T_\text{II}$ could receive significant contribution when $m_{h_S}$ is away from 125 GeV, which is negative (positive) for $m_{h_S}> (<)$ 125 GeV.  Meanwhile, the masses of $H$, $A$ or $H^\pm$ can still contribute via $T_0$ and $S_0$.

% Firstly, we can set the degenerated masses ($\Delta m_A = \Delta m_H =0$) to study the individual impact of $\alpha_{hS}$, which means that the doublet sector is on the $STU$ allowed region of 2HDM and $T_0=0$. 
In the left panel of Fig.~\ref{fig:mhsahs}, we show the 95\% C.L. allowed region from the $STU$ constraints in  $m_{h_S}$ vs $\alpha_{hS}$ plane for different values of  $\Delta m_H = \Delta m_A$.   For $\Delta m_H = \Delta m_A=0$ (blue), all values of $\alpha_{hS}$ are allowed for $m_{h_S}=m_h \approx 125$~GeV.  
The allowed region for $\alpha_{hS}$ reduces for $m_{h_S}$ away from 125 GeV: 
$|\alpha_{hS}|\lesssim 0.7$ for light $m_{h_S}=10$~GeV and $|\alpha_{hS}|\lesssim 0.2$ for $m_{h_S}=1$~TeV.

 For  $\Delta m_H = \Delta m_A=50$ GeV (orange), the 95\% C.L. allowed region shifted to right of $m_{h_S}=125$ GeV, due to the opposite signs of $T_0$ and $\Delta T_\text{II}$ for $m_{h_S}>125$ GeV. For $\Delta m_H = \Delta m_A=100$ GeV (green), $T_0$ is so large that only two thin branches in $m_{h_S}>240$ GeV and $0.5<|\alpha_{hS}|<\pi/2$ are allowed.

In the right panel of  Fig.~\ref{fig:mhsahs}, we show the 95\% C.L. $STU$ allowed region in $m_{h_S}$ vs $\Delta m_{H,A}$ plane for $\alpha_{hS}=0$ (blue), $\pi/4$(orange) and $\pi/2$ (green).  For $\alpha_{hS}=0$, 
the bound of $|\Delta m_{H}|=|\Delta m_A |\lesssim$ 80~GeV is independent of 
$m_{h_S}$.  For non-zero $\alpha_{hS}$, the allowed value in $|\Delta m_{H,A}|$ reduces for $m_{h_S}<125$ GeV, while increases for $m_{h_S}>125$ GeV. Note that all curves cross at  $m_{h_S}=125$ GeV since $\Delta T_{\rm II}$ and $\Delta S_{\rm II}$ vanish at $m_{h_S}=125$ GeV regardless of the value of $\alpha_{hS}$. There is a slight asymmetry between the positive and negative values of $\Delta m_H=\Delta m_A$. This is because  $S_0$ observable is not symmetric between positive and negative $\Delta m_{H,A}$.  $\Delta S_\text{II}$ is always positive for $m_{h_S}> m_{h_{125}}$ while the sign of  $S_0$ flips for different sign of $\Delta m_{H,A}$. Therefore, the $S$ observable is larger for positive $\Delta m_{H,A}$ and the constraint would be stronger, which leads to slightly smaller allowed region for $\Delta m_{H,A}>0$ comparing to the negative mass difference case.

\subsection{Case-III: $\alpha_{HS}\neq0$}
{The Case-III corresponds to $\alpha_{HS}\neq0$ and $c_{\beta-\alpha}=\alpha_{hS}=\alpha_{AS}=0$, when $h_S$ mixes with the doublet-like CP-even Higgs $H$. The non-zero trilinear Higgs to gauge-boson couplings include
\begin{equation}
    c_{Ah_S Z},~ c_{h_S H^\pm W^\mp},
\end{equation}
in addition to those in Eq.~(\ref{eq:case0coupling}).
The $h_S VV$ coupling, however, remains to be zero in this case.  While the additional contribution to the $S$ observable is small, the $T$ observable could receive significant contributions: 
\begin{align}
% \begin{split}
        % T \propto& c_{\alpha_{HS}}^2 [F(m_{H^\pm}^2, m_H^2) - F(m_{A}^2, m_H^2) ] + s^2_{\alpha_{HS}} [ F(m_{H^\pm}^2, m_{h_S}^2) - F(m_{A}^2, m_{h_S}^2) ] + F(m_{H^\pm}^2, m_A^2), \\
        T=&\frac{1}{16\pi s_W^2 m_W^2} [c_{\alpha_{HS}}^2 F(m_{H^\pm}^2, m_H^2) +s^2_{\alpha_{HS}}F(m_{H^\pm}^2, m_{h_S}^2) ]\notag\\& + F(m_{H^\pm}^2, m_A^2) - [c_{\alpha_{HS}}^2 F(m_{A}^2, m_H^2)  +  s^2_{\alpha_{HS}} F(m_{A}^2, m_{h_S}^2) ] .\label{eq:ta2}\\
        = &T_0 + \Delta T_\text{III},\\
        \Delta T_\text{III} &= \frac{ s_{\alpha_{HS}}^2}{16\pi s_W^2 m_W^2}[ F(m_{H^\pm}^2, m_{h_S}^2) - F(m_{A}^2, m_{h_S}^2)- F(m_{H^\pm}^2, m_{H}^2)  + F(m_{A}^2, m_{H}^2)  ].\label{eq:delt3}
% \end{split}
\end{align}
In addition to $\Delta m_{H,A}$, $\Delta m_{h_S}=m_{h_S}-m_{H^\pm}$ also enters.   

There is a numerical approximation for the $F$ function in Eq.~\eqref{eq:fij}:
% \begin{multline}
\begin{equation}
\begin{split}
&c_{\alpha}^2 [F(J^2,I^2) - F(K^2,I^2)]  + s_{\alpha}^2  [F(J^2,L^2)-F(K^2,L^2)]   \\ \approx &F( J^2, [c_\alpha^2 I + s_\alpha^2 L]^2) -F(K^2,  [c_\alpha^2 I + s_\alpha^2 L]^2).
\end{split}
        % c_{\alpha_{HS}}^2 [F(m_{H^\pm}^2, m_H^2) - F(m_{A}^2, m_H^2) ] + s^2_{\alpha_{HS}} [ F(m_{H^\pm}^2, m_{h_S}^2) - F(m_{A}^2, m_{h_S}^2) ]\\\approx
        % F(m_{H^\pm}^2, (c^2_{\alpha_{HS}} m_H+ s^2_{\alpha_{HS}}m_{h_S})^2) - F(m_{A}^2, (c^2_{\alpha_{HS}} m_H+ s^2_{\alpha_{HS}}m_{h_S})^2).
        \label{eq:approx}
        \end{equation}
% \end{multline}
Therefore, the $T$ observable can be approximated as 
\begin{equation}
T\approx \frac{ 1}{16\pi s_W^2 m_W^2} [F(m_{H^\pm}^2,  (c_{\alpha_{HS}}^2 m_H + s^2_{\alpha_{HS}}m_{h_S})^2) -  F(m_{A}^2, (c_{\alpha_{HS}}^2 m_H + s^2_{\alpha_{HS}} m_{h_S})^2) + F(m_{H^\pm}^2, m_A^2) ]
\end{equation}
which vanishes for 
\begin{equation}
    c^2_{\alpha_{HS}} m_H+ s^2_{\alpha_{HS}}m_{h_S} =  m_{H^\pm}\label{eq:Hhs},\ \ \ {\rm or\ } m_A=m_{H^\pm}.
\end{equation}

\begin{figure}[h]
\begin{center}
  \includegraphics[width=.7\linewidth]{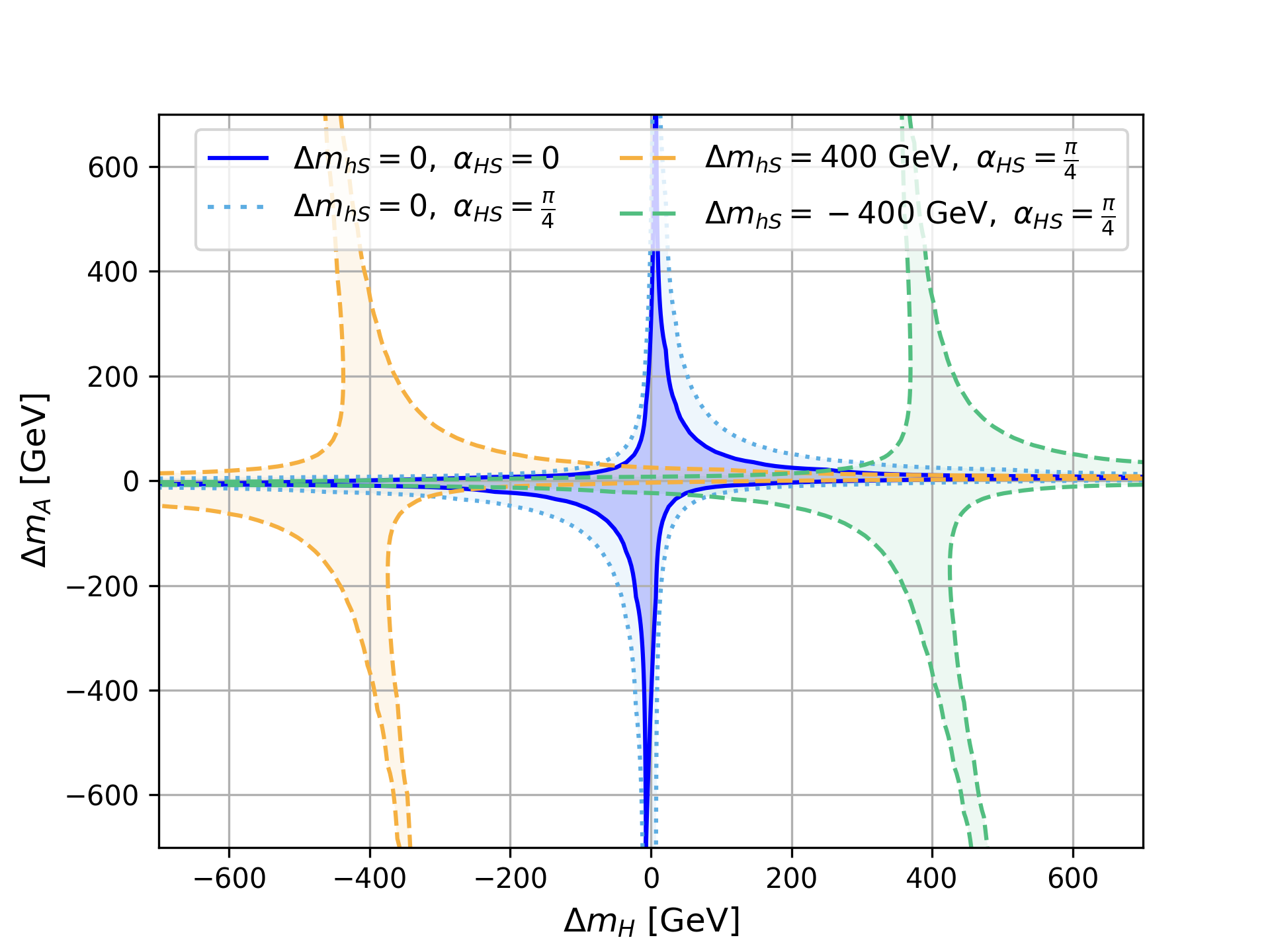}
    \caption{The 95\% C.L. $STU$ allowed region in $\Delta m_H$ vs. $\Delta m_A$ in the Case-III with $c_{\beta-\alpha}=\alpha_{hS}=\alpha_{AS}=0$.  $m_{H^\pm}$ is set to be 800 GeV. Regions enclosed by the dark solid blue and light dashed blue curves are for $\Delta m_{hS}=0$ and $\alpha_{HS}=0$ and $\pi/4$ respectively.  Orange and green regions are for $\Delta m_{hS}=\pm$ 400 GeV, respectively, and $\alpha_{HS}=\pi/4$. }
\label{fig:Case3_1}
\end{center}
\end{figure}

\begin{figure}
\includegraphics[width=.5\linewidth]{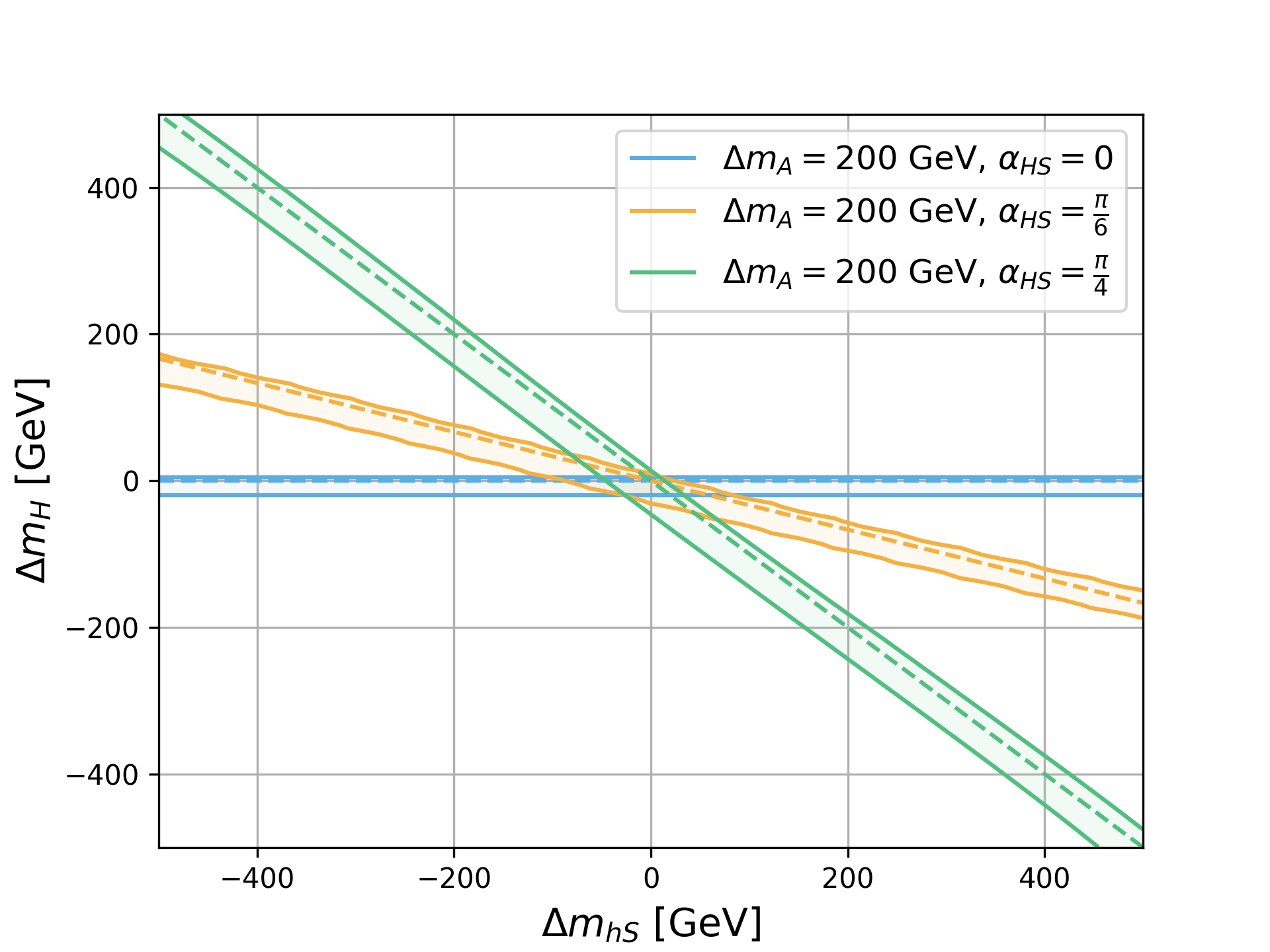}
\includegraphics[width=.5\linewidth]{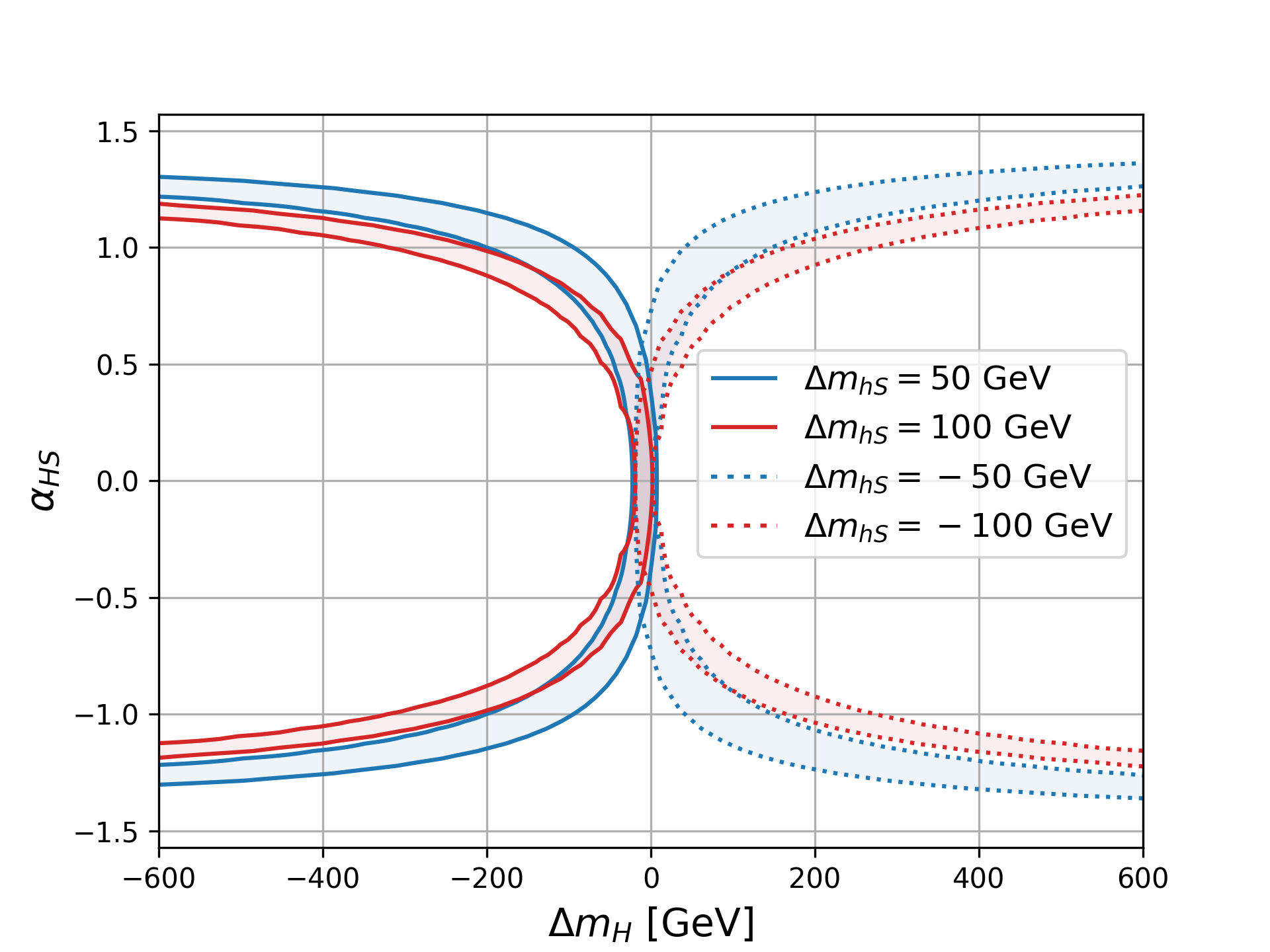}
    \caption{The 95\% C.L. $STU$ allowed region in $\Delta m_H$ vs. $\Delta m_{hS}$ plane (left panel) and  $\Delta m_{H}$ vs. $\alpha_{HS}$ plane (right panel) in the Case-III with $c_{\beta-\alpha}=\alpha_{hS}=\alpha_{AS}=0$.  In the left panel, $\alpha_{HS}$ is varied to be 0 (blue), $\pi/6$ (orange) and $\pi/4$ (green).   In the right panel, $\Delta m_{hS}$ is varied to be $\pm 50$ GeV (solid and dashed blue) and $\pm 50$ GeV (solid and dashed red).  $m_{H^\pm}$ is set to be 800 GeV and $\Delta m_A$ is set to be 200 GeV.}
\label{fig:Case3_2}  

\end{figure}

Fig.~\ref{fig:Case3_1} presents the 95\% $STU$ allowed region in the $\Delta m_H$ vs $\Delta m_A$ plane for the Case-III. The region enclosed by the dark blue curves corresponds to the baseline  Case-0 when $\alpha_{HS}=0$.  For $\alpha_{HS}=\pi/4$ and $\Delta m_{hS}=0$ (region enclosed by the light dotted blue lines), the 95\% C.L. $STU$ allowed region would be slightly enlarged compared to the Case-0
since the mass-splitting effect of $H$ with $H^\pm$ is suppressed by $c^2_{\alpha_{HS}}=1/2$, while the $h_S$ has no mass splitting with the charged Higgs, as shown in Eq.~(\ref{eq:Hhs}).

When the singlet-like Higgs mass is deviated from the charged Higgs mass, for instance, $\Delta m_{hS}=\pm 400$~GeV with $\alpha_{HS}=\frac{\pi}{4}$, as shown by the orange and green regions, the central of the allowed region in $\Delta m_H$ would be shifted to the region of $\Delta m_H\approx \mp 400$~GeV to satisfy the mass relation in Eq.~(\ref{eq:Hhs}) in order to suppress the contributions to the $T$ parameter.  Note that $\Delta m_A=0$ is still allowed, regardless of the choices of $\Delta m_H$, $\Delta m_{h_S}$ and $\alpha_{HS}$. 

In the left panel of Fig.~\ref{fig:Case3_2}, we show the 95\% C.L. $STU$ allowed region in $\Delta m_H$ vs. $\Delta m_{hS}$ plane in the Case-III  for  $m_{H^\pm}=800$ GeV and $\Delta m_A=200$ GeV, with varying $\alpha_{HS}=0$ (blue), $\pi/6$ (orange), and $\pi/4$ (green).  Also shown in dashed lines are the approximate relation of $c^2_{\alpha_{HS}}\Delta m_H=-s^2_{\alpha_{HS}} \Delta m_{hS}$ based on Eq.~(\ref{eq:Hhs}). As the mass splitting $\Delta m_A$ increases, the $STU$ bands would shrink and be closer to the dashed lines. 

In the right panel of Fig.~\ref{fig:Case3_2}, we show the 95\% C.L. $STU$ allowed region in $\Delta m_H$ vs. $\alpha_{HS}$ plane in the Case-III  for  $m_{H^\pm}=800$ GeV and $\Delta m_A=200$, with varying $\Delta m_{hS}=\pm 50$ GeV (blue) and 100 GeV (red). Note that the allowed regions are symmetric in $\alpha_{HS}$ and only have slight variation with respect to the sign of $\Delta m_H$.

\subsection{Case-IV: $\alpha_{AS}\neq0$ }
%\begin{figure}[h]
%    \centering
%     \includegraphics[width=.5\linewidth]{figs/IV/stu12.png}
%    \caption{The $STU$ constraints on the parameter space of $\Delta m_H$, $\Delta m_A$, $\Delta m_{AS}$ and $\alpha_{AS}$. All of these parameter spaces are imposed the limits of $c_{\beta-\alpha}=\alpha_{hS}=\alpha_{HS}=0$.}
%    \label{fig:case4}
%\end{figure}

In the Case-IV, ($\alpha_{AS}\neq0$ and $c_{\beta-\alpha}=\alpha_{hS}=\alpha_{HS}=0$), the CP-odd sector has singlet admixture and CP-even sector is the same as the Case-0. 
The non-zero trilinear Higgs to gauge-boson couplings include
\begin{equation}
~c_{A_S H Z},~ c_{A_S H^\pm W^\mp},
\end{equation}
in addition to those in  Eq.~(\ref{eq:case0coupling}).
Consequently, the couplings involving CP-odd Higgses are parametrized by $\alpha_{AS}$, i.e. $AHZ$ and $AH^\pm W^\mp$ depend on $c_{\alpha_{AS}}$, $A_S HZ$ and $A_S H^\pm W^\mp$ depend on $s_{\alpha_{AS}}$. The singlet CP-odd Higgs mass $m_{A_S}$ enters while the CP-even ${h_S}$ is completely decoupled.
In particular, the contributions to the $T$ observable is given by
\begin{align}
% \begin{split}
                T=&\frac{1}{16\pi s_W^2 m_W^2} [c_{\alpha_{AS}}^2 F(m_{H^\pm}^2, m_A^2) +s^2_{\alpha_{AS}}F(m_{H^\pm}^2, m_{A_S}^2) ]\notag\\& + F(m_{H^\pm}^2, m_H^2) - [c_{\alpha_{AS}}^2 F(m_{H}^2, m_A^2)  +  s^2_{\alpha_{AS}} F(m_{H}^2, m_{A_S}^2) ] ,\\
                =& T_0 + \Delta T_\mathrm{IV}\\
                \Delta T_\text{IV} &= \frac{ s_{\alpha_{AS}}^2}{16\pi s_W^2 m_W^2}[ F(m_{H^\pm}^2, m_{A_S}^2) - F(m_{H}^2, m_{A_S}^2)- F(m_{H^\pm}^2, m_{A}^2)  + F(m_{A}^2, m_{H}^2)  ]
% \end{split}
\end{align}
Comparing to Eqs.~(\ref{eq:ta2}) and (\ref{eq:delt3}), we see that the Case-IV is very similar to the Case-III, with the substitution of $H$ and $h_S$ with $A$ and $A_S$, as well as the corresponding mass parameters and mixing angles.  
The approximate expression for $T$ is
\begin{equation}
T\approx \frac{ 1}{16\pi s_W^2 m_W^2} [F(m_{H^\pm}^2,  (c_{\alpha_{AS}}^2 m_A + s^2_{\alpha_{AS}}m_{A_S})^2) -  F(m_{H}^2, (c_{\alpha_{AS}}^2 m_A + s^2_{\alpha_{AS}} m_{A_S})^2) + F(m_{H^\pm}^2, m_H^2) ],
\end{equation}
which leads to a similar approximate mass relation that satisfies the $STU$ constraints: 
\begin{align}
    c^2_{\alpha_{AS}} m_A + s^2_{\alpha_{AS}}m_{A_S} = m_{H^\pm}, \ \ \ {\rm or\ } m_H=m_{H^\pm}.\label{eq:Aas}
\end{align}
The 95\% C.L. $STU$ allowed regions in $\Delta m_H$ vs. $\Delta m_A$, $\Delta m_{AS}$ vs. $\Delta m_A$, and $\Delta m_A$ vs. $\alpha_{AS}$ are very similar to those presented in Fig.~\ref{fig:Case3_1}$-$Fig.~\ref{fig:Case3_2}, with the switching of $A\leftrightarrow H$.

In general, the mass splittings of $\Delta m_H$ and $\Delta m_A$ can contribute the $STU$ observables via the $AHZ$, $AH^\pm W^\mp$ and $H H^\pm W^\mp$ loops. In the 2HDM+S, the singlet CP-even Higgs $h_S$ can mix with $H$ via the mixing $\alpha_{HS}$, and the singlet CP-odd Higgs $A_S$ can mix with $A_S$ via $\alpha_{AS}$. Therefore, $\Delta m_{hS}$, $\Delta m_{AS}$, as well as  $\alpha_{HS}$ and $\alpha_{AS}$ enter. The $STU$ constraints can be still fulfilled when the mass relations in Eqs.~\eqref{eq:Hhs} or \eqref{eq:Aas} are satisfied.

% However, as Figs.~\ref{fig:hhs-aas} show, the minimal $chi^2$ lines are both apart from the dahsed lines for larger mass splittings, where the dashed lines represent the lines satisfying the Eqs.~\eqref{eq:Hhs} and \eqref{eq:Aas}. This means these relations are not the exact equations, and Eq.~\eqref{eq:approx} is only the approximation form. Therefore, Eqs.~\eqref{eq:Hhs} and \eqref{eq:Aas} cannot lead to the minimum $\chi^2$ of $STU$ fit. Nevertheless, the $STU$ allowed parameter spaces are still around regions where the Eqs.~\eqref{eq:Hhs} and \eqref{eq:Aas} are fulfilled.

\section{The $STU$ constraints beyond the alignment limit}
\label{sec:res2}

For the Case-I, we consider the non-alignment limit with all the single mixing angles set to be zero.  For the Case-II $-$ IV, we focus on the scenario with only one mixing angle is set to be non-zero under the alignment limit. In this section, we consider the cases with a non-zero  singlet mixing angle beyond the alignment limit.

\begin{figure}[h]
\centering
% \subfigure[]{\label{fig:cbamhsahs}
\includegraphics[width=0.5\linewidth]{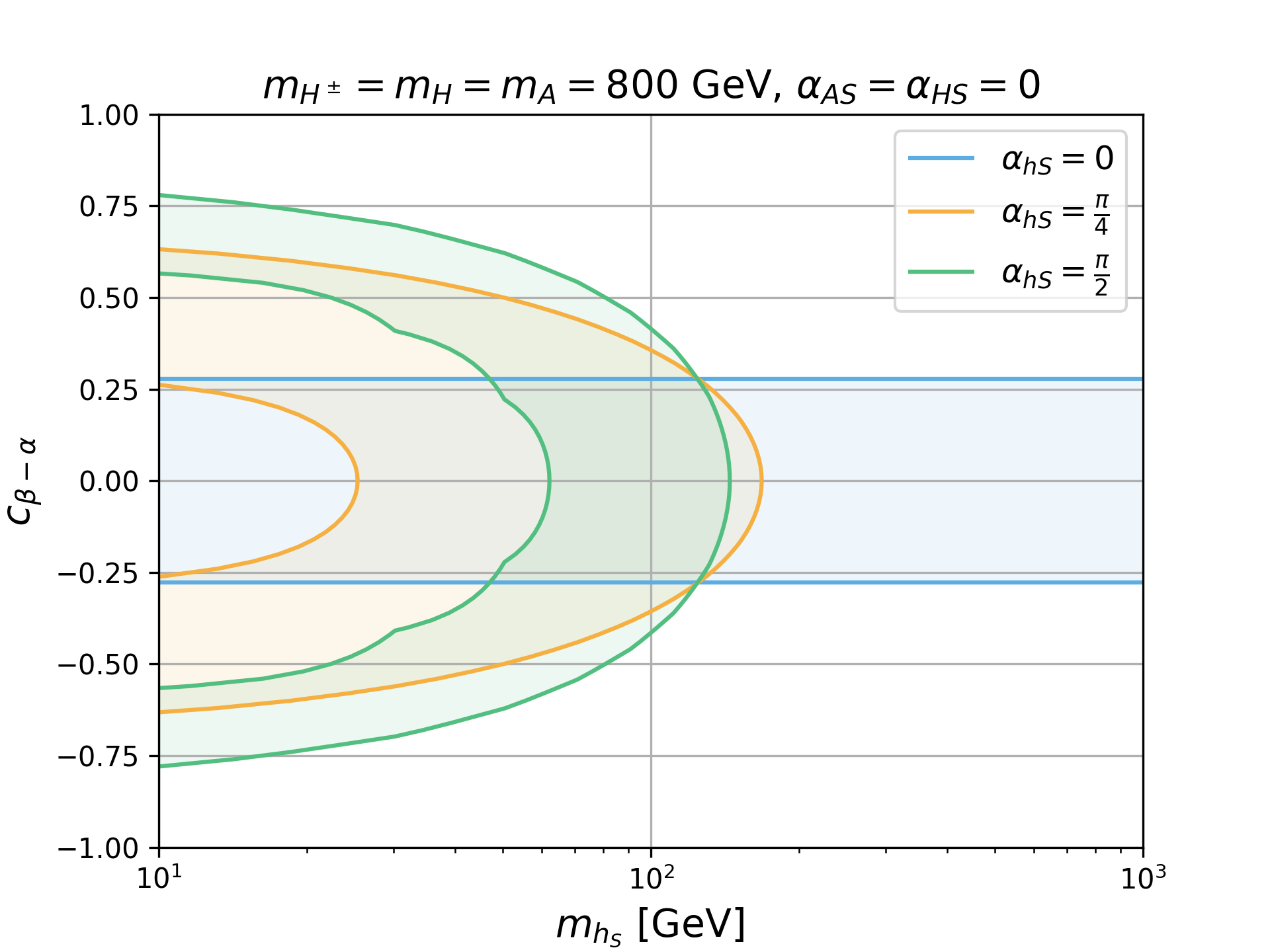}\includegraphics[width=0.5\linewidth]{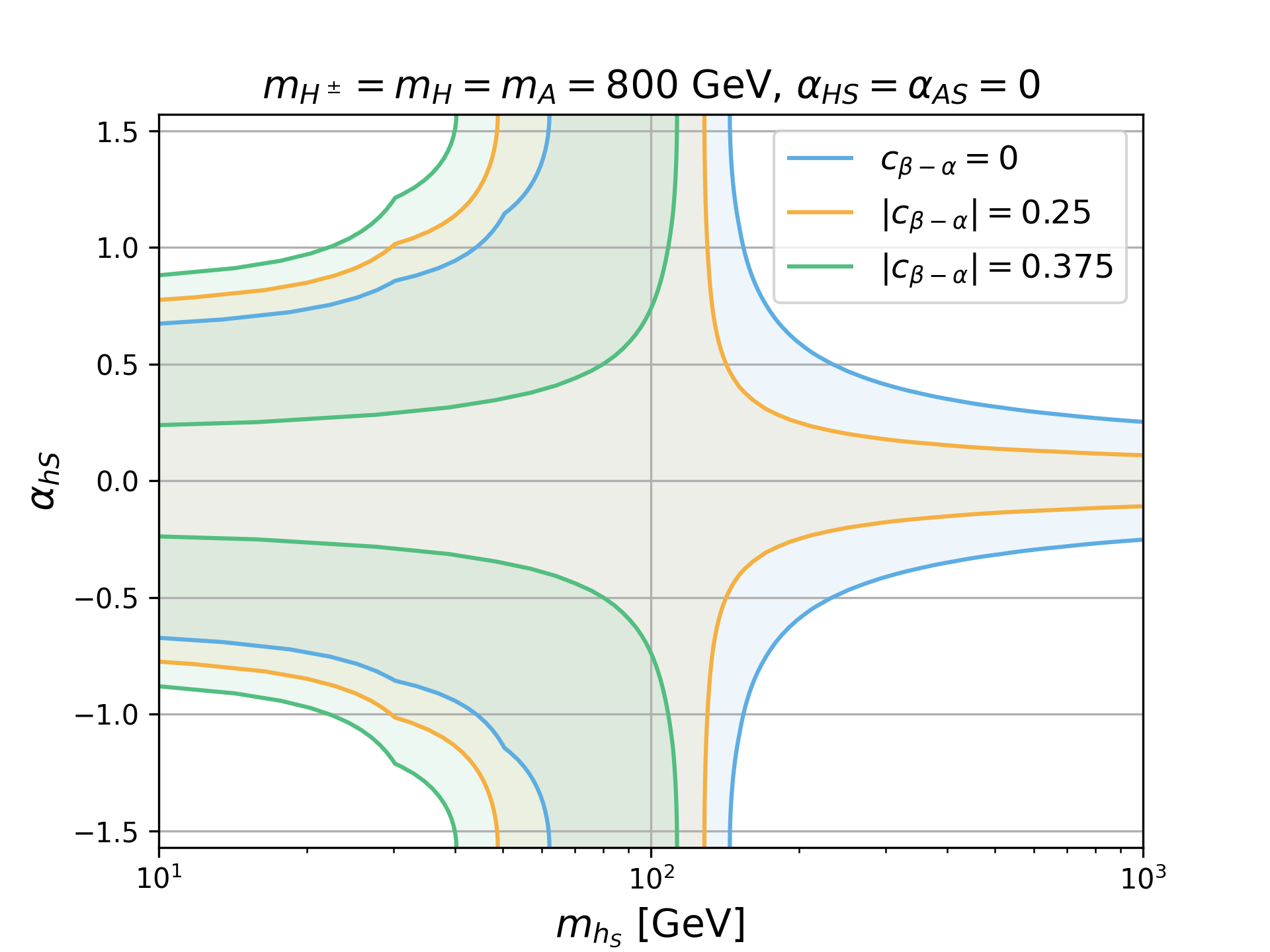}
% }
% \subfigure[]{\label{fig:ahsmhscba}
% \includegraphics[width=0.5\linewidth]{figs/beyond/ahsmhs.png}
% }
\caption{95\% C.L. $STU$ allowed region in $m_{h_S}$ vs. $c_{\beta-\alpha}$ plane (left panel) for various $\alpha_{hS}=0$ (blue), $\pi/4$ (orange)  $\pi/2$ (green), and $m_{h_S}$ vs. $\alpha_{hS}$ plane (right panel) for various $c_{\beta-\alpha}=0$ (blue), 0.25 (orange) and 0.375 (green).  We set $m_{H^\pm}=m_{H}=m_A=800$ GeV and $\alpha_{AS}=\alpha_{HS}=0$.  }
\label{fig:ahsmhscba}
% \Wei{$\cos_{\beta-\alpha}$,$c_{\beta-\alpha}$? sometimes we also have $\cos(\beta-\alpha)$}
\end{figure}
We first explore the interplay between $c_{\beta-\alpha}$ with the singlet$-$$h_{125}$ mixing $\alpha_{hS}$. In the left plot of Fig.~\ref{fig:ahsmhscba}, we show the 95\% C.L. $STU$ allowed region in $m_{h_S}$ vs. $c_{\beta-\alpha}$ plane for various $\alpha_{hS}$. For $\alpha_{hS}=0$ (region enclosed by the solid blue curve), $|c_{\beta-\alpha}|$ is constrained to be less than 0.275, independent of $m_{h_S}$.  However, the singlet admixture can enlarge the allowed region in $|c_{\beta-\alpha}|$, as shown by the two elliptical rings for $\alpha_{hS}=\pi/4$ and $\pi/2$.  $h_SVV$ interaction can compensate the contribution of  $\Delta T_\text{I}$ in. Eq.~\eqref{eq:t11} for larger $|c_{\beta-\alpha}|$.

In the right plot of Fig.~\ref{fig:ahsmhscba}, we present the 95\% C.L. $STU$ allowed region in $m_{h_S}$ vs. $\alpha_{hS}$ plane for various $c_{\beta-\alpha}$. For increasing $|c_{\beta-\alpha}|$, the allowed region shifts to the left.  For $m_{h_S}>125$ GeV, the allowed range of $\alpha_{hS}$ reduces, while for $m_{h_S}<125$ GeV, larger values of $\alpha_{hS}$ are allowed. For $|c_{\beta-\alpha}|$ slightly above 0.275, $\alpha_{hS}=0$ is no longer allowed, and   two branches appear. 

\begin{figure}[h]
    \centering
   % \hfill\subfigure[]{\label{fig:cbamh2}
   \includegraphics[width=0.5\linewidth]{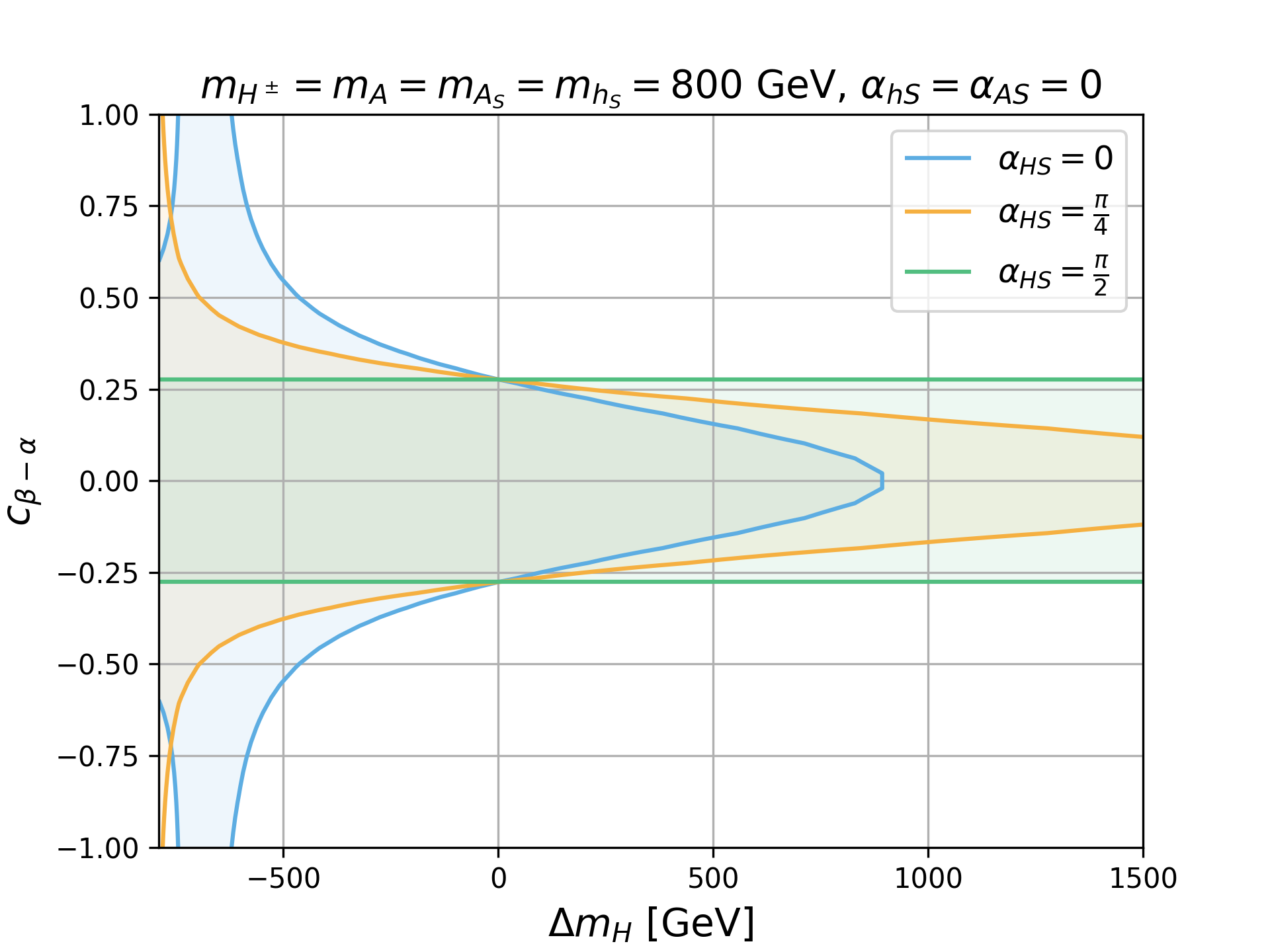}\includegraphics[width=0.5\linewidth]{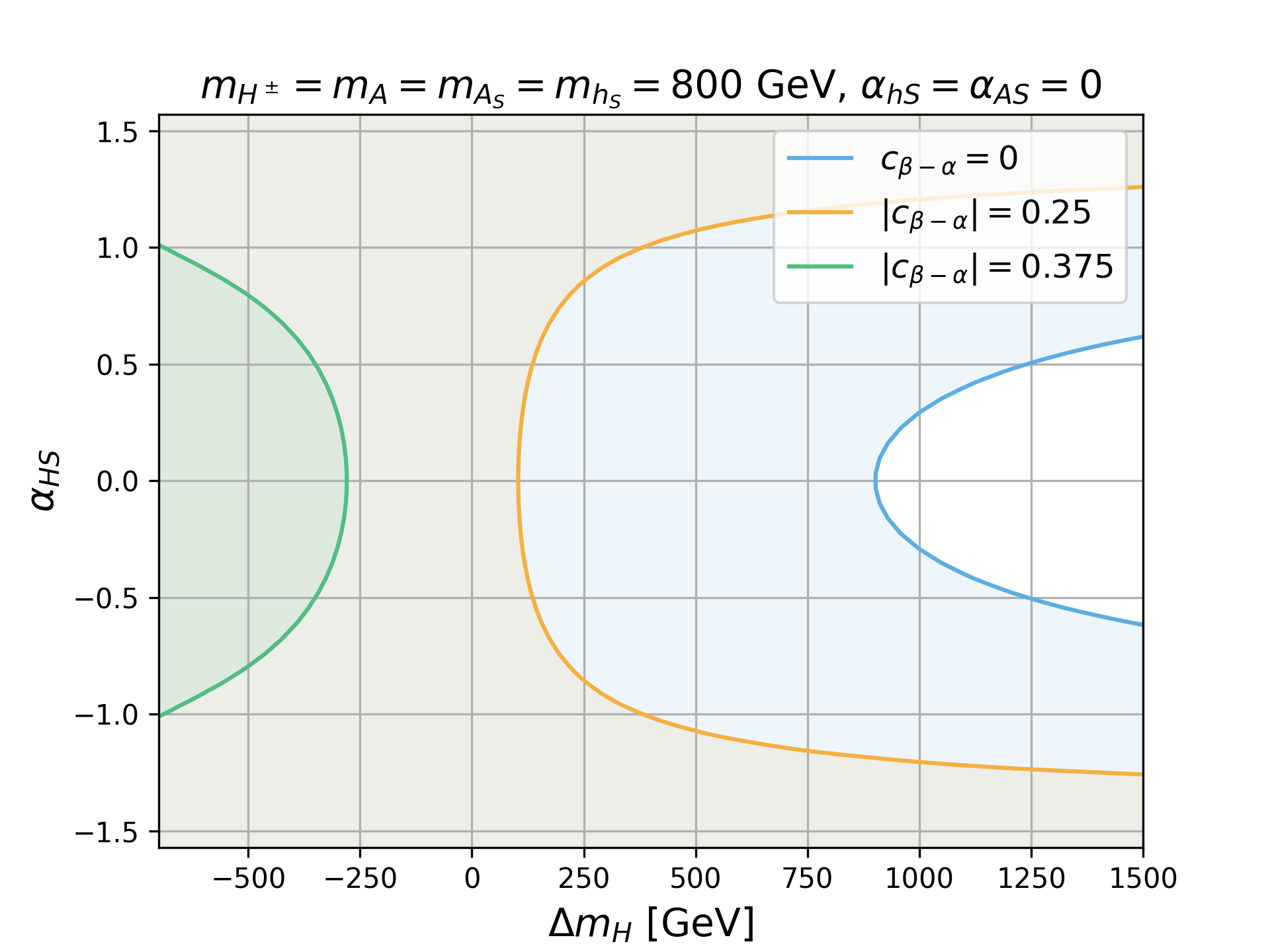}
   % }
    % \subfigure[]{\label{fig:a2mh}
    % \includegraphics[width=0.5\linewidth]{figs/beyond/a2mh.png}
    % }

    % \subfigure[]{\label{}\includegraphics[width=0.475\linewidth]{figs/beyond/cbamhs.png}}
    % \subfigure[]{\label{fig:ahsmhscb}\includegraphics[width=0.475\linewidth]{figs/beyond/a2mhs.png}}
    
    \caption{95\% C.L. $STU$ allowed region in $\Delta m_{H}$ vs. $c_{\beta-\alpha}$ plane (left panel) for various $\alpha_{HS}$=0 (blue), $\pi/4$ (orange),  $\pi/2$ (green), and $\Delta m_{H}$ vs. $\alpha_{HS}$ plane (right panel) for various $|c_{\beta-\alpha}|$=0 (blue), 0.25 (orange) and 0.375 (green). We set $m_{H^\pm}=m_A=m_{h_S}=m_{A_S}=800$ GeV and $\alpha_{AS}=\alpha_{hS}=0$.  }
        \label{fig:cbamha2}
\end{figure}

We explore the interplay between $c_{\beta-\alpha}$ and singlet-double CP-even Higgs $H$ mixing $\alpha_{HS}$ in Fig.~\ref{fig:cbamha2}.  Left Panel show  the 95\% C.L. $STU$ 
 allowed region in $\Delta m_{H}$ vs. $c_{\beta-\alpha}$ plane for $\alpha_{HS}=0$ (blue), $\pi/4$ (orange), and $\pi/2$ (green).  The   blue line in the left panel of Fig.~\ref{fig:cbamha2} is consistent with the blue curve in the left panel of Fig.~\ref{fig:2hdmcase} (Case-I).  For larger $\alpha_{HS}$, the allowed range of $c_{\beta -\alpha}$ shrinks for $\Delta m_H < 0$ while expands for $\Delta m_H > 0$. In this case, $c_{HVV} = c_{\beta-\alpha}c_{\alpha_{HS}}$, $c_{hVV} = s_{\beta-\alpha}$ and $c_{h_SVV} = -c_{\beta-\alpha}s_{\alpha_{HS}}$. As $\alpha_{HS}$ increases, the $HVV$ contribution to $T$ observable decreases, while $h_SVV$ contribution increases. Therefore, the allowed $c_{\beta-\alpha}$ region shrinks in the $h_SVV$ dominate region ($m_H<m_{h_S}$) and enlarges in the $HVV$ dominate region ($m_H>m_{h_S}$). When $m_H=m_{h_S}=$800~GeV, the $h_SVV$ term plays the same role as $HVV$. The $STU $ constraints of this point are independent of $\alpha_{HS}$ and all curves cross at $\Delta m_H=0$. 
For $\alpha_{HS}=\frac{\pi}{2}$, $H$ becomes the pure singlet Higgs and does not contribute to the $STU$ parameters. Therefore, the $STU$ limit of $|c_{\beta-\alpha}|<0.275$ is independent of $m_H$. 
%\Shufang{ Test mHpm=1 TeV. Is the crossing point still 0.25?}
%This limit of $|c_{\beta-\alpha}|$ would slightly decrease as $m_{H^\pm}=m_H=m_{h_S}=m_A=m_{A_S}$ both increase. 
Since the role of $H$ and $h_S$ switches when $\alpha_{HS} \rightarrow \pi/2-\alpha_{HS}$, the parameter space of $c_{\beta-\alpha}$ vs $\Delta m_{h_S}$ is the same as the left panel of Fig.~\ref{fig:cbamha2} with  $\alpha_{HS}\rightarrow \pi/2-\alpha_{HS}$.

The right panel of Fig.~\ref{fig:cbamha2} presents the  $\Delta m_{H}$ vs. $\alpha_{HS}$ plane for $|c_{\beta-\alpha}|$=0 (blue), 0.25 (orange) and 0.375 (green). For  $c_{\beta-\alpha}=0$, almost the entire region of parameter space is allowed, except for a small open region with relatively small $\alpha_{HS}$ and large $\Delta m_H$.  The allowed region reduces when $|c_{\beta-\alpha}|$ increases.  For $|c_{\beta-\alpha}|=0.375$, only a small region with $\Delta m_H<-250$ GeV and $|\alpha_{HS}|<1$ is allowed. This is due to the increased contribution from $HVV$ term at larger $|c_{\beta-\alpha}|=0.375$. Only when the $m_H$ is lighter and close to 125~GeV, the $HVV$ contribution would be small enough to be allowed. %Particularly, the $\alpha_{HS}=0$ would be consistent with $\alpha_{HS}=0$ in Fig.~\ref{fig:2hdmcase}, and the larger singlet admixture would suppress the limit of $m_H$, since $m_{h_S}=800$~GeV is much heavier.
Similar to the left panel, the parameter space of $\alpha_{HS}$ vs $\Delta m_{h_S}$ is the same as the right panel of Fig.~\ref{fig:cbamha2} with $\alpha_{HS}\rightarrow \pi/2-\alpha_{HS}$.

\begin{figure}[h]
    \centering
   % \hfill\subfigure[The parameter space of $c_{\beta-\alpha}$ vs $m_A$ with different $\alpha_{AS}$.]{\label{fig:cbama2}
   \includegraphics[width=0.5\linewidth]{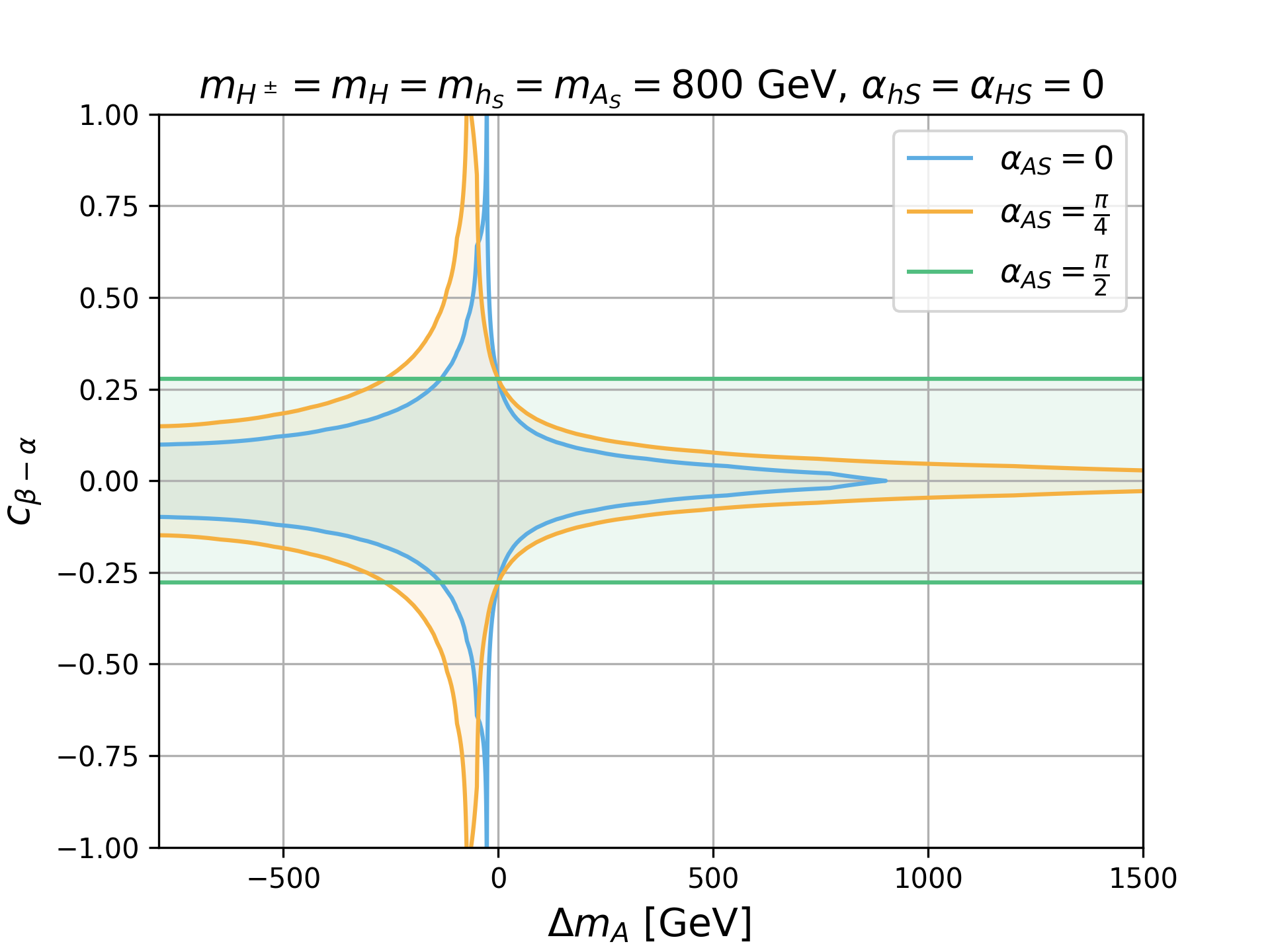}\includegraphics[width=0.5\linewidth]{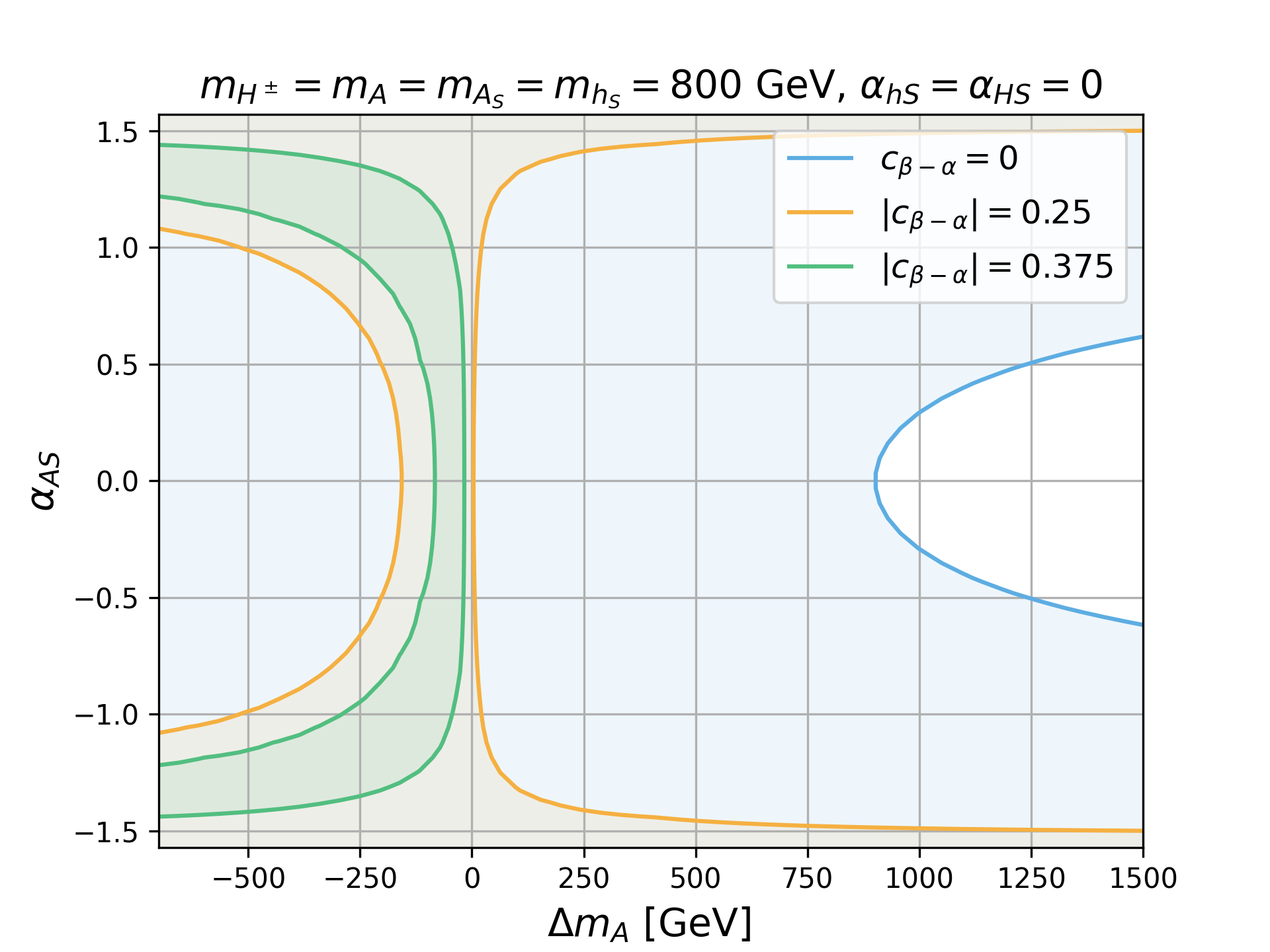}%}
    % \subfigure[The parameter space of $\alpha_{AS}$ vs $\Delta m_A$ with different $c_{\beta-\alpha}$.]{\label{fig:a4ma}
    % \includegraphics[width=0.5\linewidth]{figs/beyond/a4ma.png}%}
        % \subfigure[]{\label{}\includegraphics[width=0.475\linewidth]{figs/beyond/cbamas.png}}\subfigure[]{\label{fig:asmascb}\includegraphics[width=0.475\linewidth]{figs/beyond/aasmas.png}}
    \caption{95\% C.L. $STU$ allowed region in $\Delta m_{A}$ vs. $c_{\beta-\alpha}$ plane (left panel) for various $\alpha_{HS}$=0 (blue), $\pi/4$ (orange),  $\pi/2$ (green), and $\Delta m_{A}$ vs. $\alpha_{AS}$ plane (right panel) for various $|c_{\beta-\alpha}|$=0 (blue), 0.25 (orange) and 0.375 (green). We set $m_{H^\pm}=m_H=m_{h_S}=m_{A_S}=800$ GeV and $\alpha_{hS}=\alpha_{HS}=0$.}
    \label{fig:cbamaa4}
\end{figure}

We explore the interplay between $c_{\beta-\alpha}$ and singlet-double CP-odd Higgs $A$ mixing $\alpha_{AS}$ in Fig.~\ref{fig:cbamaa4}. The left panel presents the 95\% C.L. allowed region in $\Delta m_A$ vs. $c_{\beta-\alpha}$ for various $\alpha_{HS}$=0 (blue), $\pi/4$ (orange),  and $\pi/2$ (green).  The blue region in the left panel of Fig.~\ref{fig:cbamaa4} for $\alpha_{AS}=0$ is consistent with the blue region in  the right panel of Fig.~\ref{fig:2hdmcase}.
For larger $\alpha_{AS}$, the allowed regions shift to the left, whereas the $c_{\beta-\alpha}$ bounds at both $m_{A}>m_{A_S}$ and $m_{A}<m_{A_S}$ become larger, due to the suppression of both the $A H Z$ and $AhZ$ terms by $c_{\alpha_{AS}}$. However, the $A_S hZ $ term is enhanced by $s_{\alpha_{AS}}$, which compensates the suppression of $A H Z$ and $AhZ$. Therefore, the $STU$ limit is relaxed faster at $m_{A}>m_{A_S}$ where $A_S$ is less dominant in this region. For $\alpha_{AS}=\pi/2$, only $A_S h Z$ contribution is left   and contribution from $A$ decouples. 
 The 95\% C.L. allowed region for $c_{\beta-\alpha}$ limit is a constant and  independent of $m_A$.   Since the role of $A$ and $A_S$ switches when $\alpha_{AS} \rightarrow \pi/2-\alpha_{AS}$, the parameter space of $c_{\beta-\alpha}$ vs $\Delta m_{A_S}$ is the same as the left panel of Fig.~\ref{fig:cbamaa4} with  $\alpha_{AS}\rightarrow \pi/2-\alpha_{AS}$.

The right panel of Fig.~\ref{fig:cbamaa4} presents the 95\% C.L. allowed region in $\Delta m_A$ vs. $\alpha_{AS}$ for various $|c_{\beta-\alpha}|$=0 (blue), 0.25 (orange) and 0.375 (green). The   blue line indicates the same behavior of the $STU$ dependence on $(m_A, \alpha_{AS})$ as $(m_H, \alpha_{HS})$ for $c_{\beta-\alpha}=0$. However, these two cases differ when singlet admixture enters for $c_{\beta-\alpha} \neq 0$. The CP-odd Higgs $A$ enters via $A h Z$ and $A HZ$s, where these contributions are suppressed when $m_{A}$ is closed to $m_H$. In the case that $m_{A_S}=m_H$, the allowed regions only show up in the region of large $\alpha_{AS}$ are non-zero $\Delta m_A$, since $A_S$ in this area is already dominated by the doublet properties. Similar to the left panel, the parameter space of $\alpha_{AS}$ vs $\Delta m_{A_S}$ is the same as the right panel of Fig.~\ref{fig:cbamaa4} with $\alpha_{AS}\rightarrow \pi/2-\alpha_{AS}$.

\section{Interplay of electroweak and Higgs precision measurements}
\label{sec:Hprecision}

\begin{figure}[h]
    \centering
    % \subfigure[The type-I 2HDM+S]{\label{fig:h125pstyp1}
    \includegraphics[width=0.5\linewidth]{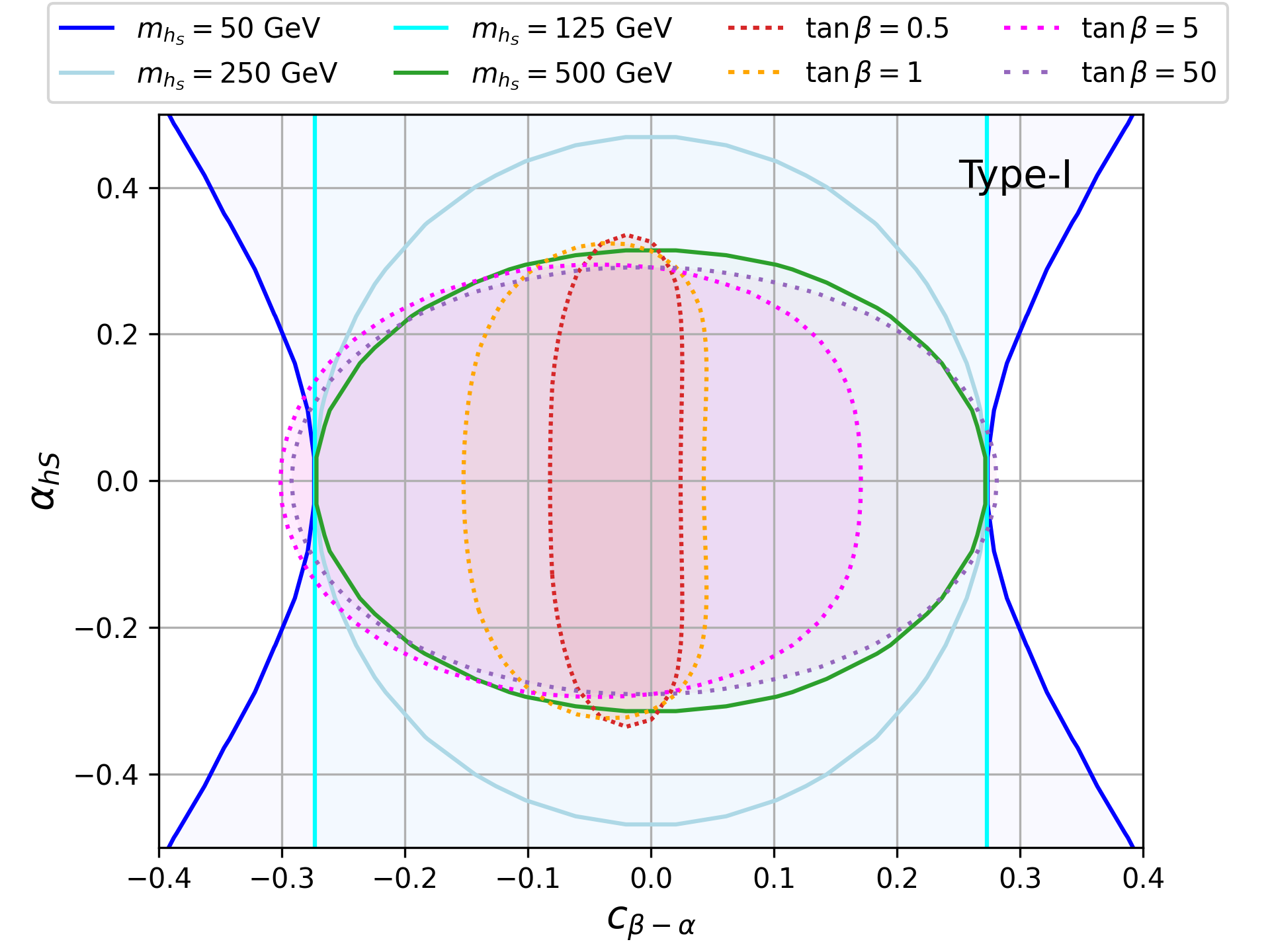}\includegraphics[width=0.5\linewidth]{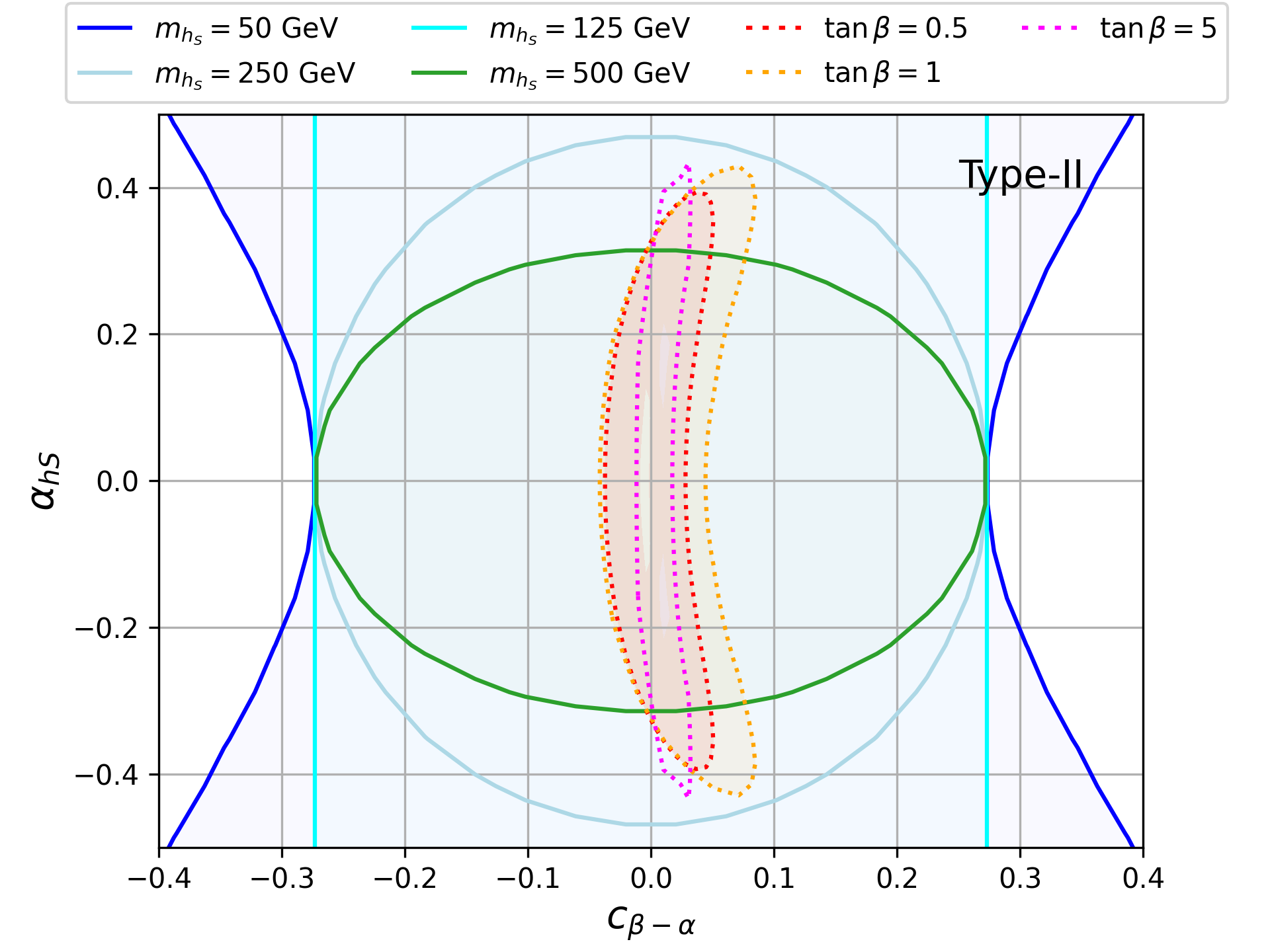}
    % }\hfill\subfigure[The type-II 2HDM+S]{\label{fig:h125pstyp2}
 %}%\subfigure[The parameter space of $\alpha_{HS}$ vs $\alpha_{hS}$, where the mass of singlet-like Higgs $m_{h_S}$ is fixed to 95~GeV. The other Higgs masses are $m_A = m_H=m_{H^\pm}=800$~GeV, and the singlet CP-odd Higgs is decoupled by $\alpha_{AS}=0$. The orange region corresponds to the $1\sigma$ region of 95~GeV excess~\cite{Biekotter:2023oen}.]{\label{fig:h95ps}\includegraphics[width=0.475\linewidth]{figs/beyond/h95.png}}
    \caption{The parameter space of $c_{\beta-\alpha}$ vs $\alpha_{hS}$, for $m_{h_S}=$50~GeV (dark blue), 125~GeV (cyan), 250~GeV (light blue), 500~GeV (green), and  $\tan\beta=$0.5 (red), 1 (orange), 5 (magenta), and 50 (purple) under electroweak precision measurements (solid curves) and Higgs precision measurements (dashed curves). The other Higgs masses are $m_A = m_H=m_{H^\pm}=800$~GeV, and $\alpha_{HS}=\alpha_{AS}=0$. Left panel is for the type-I 2HDM+S and right panel is for the type-II 2HDM+S. }
    \label{fig:specscen}
\end{figure}

The precision measurements of the couplings of the 125 GeV Higgs at the LHC also place strong constraints on the parameter space of 2HDM+S, in particular, on $c_{\beta-\alpha}$, the singlet-$h_{125}$ mixing $\alpha_{hS}$, and $\tan\beta$. We perform the fit for 125~GeV Higgs properties with \texttt{HiggsTools} \cite{Bahl:2022igd,Bechtle:2013xfa,Stal:2013hwa,Bechtle:2014ewa,Bechtle:2020uwn}.
In Fig.~\ref{fig:specscen}, we present both 95\% C.L. $STU$ allowed region in $c_{\beta-\alpha}$ vs. $\alpha_{hS}$ place for various $m_{h_S}$ (regions enclosed by solid curves) and 95\% C.L. allowed region by 125 GeV Higgs precision measurements for various $\tan\beta$ (region enclosed by dashed curves) for the Type-I (left panel) and the Type-II (right panel). 
Since the $STU$ constraints only depend on the couplings of the Higgses with the gauge bosons, which is the same for different types of 2HDM, the solid curves are the same at both panels.  For $m_{h_S}=125$ GeV, the allowed range in $c_{\beta-\alpha}$ is independent of $\alpha_{hS}$.  This is because $h_SVV$ and $hVV$ contribute the same for $m_{h}=m_{h_S}$, and $\alpha_{hS}$ is not constrained as shown in the left plot of Fig.~\ref{fig:mhsahs}. For $m_{h_S}=50$ GeV, a larger region of $c_{\beta-\alpha}$ can be accommodated for $\alpha_{h_S} \neq 0$ since the larger $\Delta T_\text{I}$ can be compensated by $h_SVV$ with lighter $m_{h_S}$.  While for $m_{h_S}>125$ GeV, the allowed region in $\alpha_{h_S}$ shrinks, where the $\Delta T_\text{I}$ and $h_S VV$ have the same sign in this mass region, which leads to tighter constraints.   

For the Higgs precision on the Type-I 2HDM+S in the left panel, the allowed range of $c_{\beta-\alpha}$ gets weaker for larger $\tan\beta$.   For $\tan\beta\gtrsim5$, the electroweak precision measurements provide a stronger constraint on $c_{\beta-\alpha}$ at negative $c_{\beta-\alpha}$ region, while the Higgs precision measurements constrain the value of $\alpha_{h_S}$ better for $m_{h_S} \lesssim 500$ GeV. For $\tan\beta\sim50$, the $STU$ constraint on positive $c_{\beta-\alpha}$ can be stronger than Higgs precision measurement at $\alpha_{hS}\sim0$. 
Thus, the $|\alpha_{hS}|$ in type-I model would be constrained to be less than 0.3 by the $h_{125}$ coupling measurements, where the $STU$ can provide a stronger $\alpha_{hS}$ limit for $m_{h_S}>500$~GeV.

For the Higgs precision on the Type-II 2HDM+S in the right panel, the allowed region in $c_{\beta-\alpha}$ is constrained to be much tighter, only a thin region around $c_{\beta-\alpha}\sim 0$. The constraints from the 
$h_{125}$ coupling measurements are the weakest at $\tan\beta\sim1$ and become stronger as  $\tan\beta$ increases or decreases.  $|\alpha_{hS}|$ is constrained to be around 0.4, which is less dependent on the values of $\tan\beta$. The electroweak precision measurements provide a tight bound on the range of $\alpha_{hS}$ for $m_{h_S} > 250$~GeV.  A combination of  the electroweak precision measurements and the Higgs precision measurements could help us to nail down the parameter space of 2HDM+S.

%%%%%%%%%%%%%% section %%%%%%%%%%%%%%%%%%%%%%%
\section{Conclusions}
\label{sec:conclu}
In this paper, we studied the implications of the oblique parameters,  in particular, the $T$ parameter, on the parameter space of the 2HDM+S model. Nine model parameters enter, including five masses $m_H$, $m_{h_S}$, $m_A$, $m_{A_S}$, $m_{H^\pm}$, and four mixing angles $c_{\beta-\alpha}$, $\alpha_{hS}$, $\alpha_{HS}$ and $\alpha_{S}$. We identified five benchmark scenarios, Case-0 with $c_{\beta-\alpha}=0$ and all the singlet mixing angles being 0 (the 2HDM alignment limit), and Case-I to IV with only one of the  mixing angles being non-zero.  We studied the 95\% C.L. $STU$ allowed region in the relevant parameter spaces. 
We fount that
\begin{itemize}
\item{\bf Case-0}\\Other than the well known conclusion that the electroweak precision constraints are satisfied for $\Delta m_{H}=0$ or $\Delta m_A=0$, there is an upper limit on the mass splitting of $\Delta m_{H/A}\lesssim$ 900~GeV for $m_{H^\pm}=800$~GeV and $\Delta m_{A,H}=0$, coming from the $S$ parameter constraint.  This upper limit also varies with $m_{H^\pm}$.
\item{\bf Case-I with $c_{\beta-\alpha} \neq 0$}\\ The constraint on $c_{\beta-\alpha}$ is weak for $m_H=125$ GeV, $\Delta m_A=0$, or $m_H=m_{H^\pm}$ and $\Delta m_A\sim -30$ GeV.  The parameter space in $\Delta m_H-c_{\beta-\alpha}$ or $\Delta m_A - c_{\beta-\alpha}$ is significantly reduced for $\Delta m_A$ or $\Delta m_H$ away from 0.
\item{\bf Case-II with $\alpha_{hS} \neq 0$ }\\ $\alpha_{hS}$ is unconstrained for $m_{h_S}=125$ GeV and $\Delta m_{H,A}=0$.  Allowed region shifts to larger $m_{h_S}$ and $|\alpha_{hS}|$ for  $\Delta m_{A,H}\neq 0$. 
\item{\bf Case-III with $\alpha_{HS} \neq 0$}\\ $STU$ constraint can be satisfied for $c^2_{\alpha_{HS}} m_H+ s^2_{\alpha_{HS}}m_{h_S} =  m_{H^\pm}$ or $m_A=m_{H^\pm}$.
\item{\bf Case-IV with $\alpha_{AS} \neq 0$}\\ $STU$ constraint can be satisfied for $c^2_{\alpha_{AS}} m_A+ s^2_{\alpha_{AS}}m_{A_S} =  m_{H^\pm}$ or $m_H=m_{H^\pm}$.
\end{itemize}

We further explored case-II - IV with non-zero $c_{\beta-\alpha}$, and observed that the singlet extension could compensate the $c_{\beta-\alpha}$  contribution and extend the allowed parameter space.  Larger $|c_{\beta-\alpha}|$, however, typically leads to more constrained mass vs. mixing angle parameter space.

We also studied the complementarity between the electroweak precision analyses and Higgs coupling measurements. We found that for the Type-I scenario, electroweak precision measurements provide stronger constraints on $\alpha_S$ for $m_{hS}>500$ GeV, while Higgs coupling measurements constrain $c_{\beta-\alpha}$ better for $\tan\beta>5$.  For the Type-II scenario, the electroweak precision measurements provide a tight bound on the range of $\alpha_{hS}$ for $m_{h_S} > 250$~GeV while the Higgs coupling measurements constrain $c_{\beta-\alpha}$ better for all values of $\tan\beta$.

In summary, the singlet extension of the 2HDM opens up the allowed parameter space when constraints from the electroweak precision measurements are considered. It also provides a complementary reach when combined with Higgs precision measurements.  Our results are independent of the detailed symmetry assumption of Higgs sector and can be applied to the general case of the 2HDM+S.  While our work  studied benchmark scenarios with only one singlet mixing angle being nonzero, it identifies the main features of each mixing case, and provides a better understanding of a more comprehensive study in the most general mixing cases.

\section*{Acknowledgments}
%%%%%%%%%%%%%%%%%%%%%%%%%%%%%%%%%%%%%%%%%%%%%%
%%%%%%%%%%%%%% subsection %%%%%%%%%%%%%%%%%%%%%%%

%%%%%%%%%%%%%%%%%%%%%%%%%%%%%%%%%%%%%%%%%%%%%%
CL, JL and WS are supported by the Natural Science Foundation of China (NSFC) under grant numbers 12305115, Shenzhen Science and Technology Program (Grant No. 202206193000001, 20220816094256002), Guangdong Provincial Key Laboratory of Gamma-Gamma Collider and Its Comprehensive Applications(2024KSYS001), and Guangdong Provincial Key Laboratory of Advanced Particle Detection Technology(2024B1212010005). JL is also supported by the Fundamental Research Funds for the Central Universities, and the Sun Yat-sen University Science Foundation. SS is supported by the Department of Energy under Grant No. DEFG02-13ER41976/DE-SC0009913.

%%%%%%%%%%%%%%%%%%%%

\appendix
\section{Appendix}
\label{sec:append}
The $STU$ observables are defined by
 \begin{align}
     &\alpha(m_Z)T = \frac{\Pi_{WW}(0)}{m_W^2}-\frac{\Pi_{ZZ}(0)}{m_Z^2},
     \label{eq:T}\\
     &\frac{\alpha(m_Z)}{4 s_W^2 c_W^2}S=\frac{\Pi_{ZZ}(m_Z^2)-\Pi_{ZZ}(0)}{m_Z^2}-\frac{c_W^2 -s_W^2}{s_W c_W}\frac{\Pi_{Z\gamma}(m_Z^2)}{m_Z^2}-\frac{\Pi_{\gamma\gamma}(m_Z^2)}{m_Z^2},
     \label{eq:S}\\ 
     &\frac{\alpha(m_Z)}{4s_W^2}(S+U)=\frac{\Pi_{WW}(m_W^2)-\Pi_{WW}(0)}{m_W^2}-\frac{c_W}{s_W}\frac{\Pi_{Z\gamma}(m_Z^2)}{m_Z^2}-\frac{\Pi_{\gamma\gamma}(m_Z^2)}{m_Z^2},
     \label{eq:SU}
 \end{align}
where $F$, $G$ and $\hat{G}$ functions are defined as~\cite{Grimus:2008nb}
\begin{equation}
    F(I,J)=\begin{cases} 
        \frac{I+J}{2}-\frac{IJ}{I-J}\ln\frac{I}{J} & \text{for } I\neq J\\
        0 & \text{for } I = J
    \end{cases}, \label{eq:fij}
\end{equation}
\begin{align}
    \begin{split}
        G(I,J,Q) &=-\frac{16}{3}+\frac{5(I+J)}{Q}-\frac{2(I-J)^2}{Q^2}+\frac{r}{Q^3}f(I+J-Q,Q^2-2Q(I+J)+(I-J)^2)\\
        &+\frac{3}{Q}\Bigg[ \frac{I^2+J^2}{I-J}-\frac{I^2-J^2}{Q}+\frac{(I-J)^3}{3Q^2}\Bigg]\ln\frac{I}{J},
    \end{split}\label{eq:Gijq}\\
    \begin{split}
        \hat{G}(I,Q) &=-\frac{79}{3}+9\frac{I}{Q}-2\frac{I^2}{Q^2}+\left(12-4\frac{I}{Q}+\frac{I^2}{Q^2}\right)\frac{f(I,I^2-4IQ)}{Q}
        \\&+\left(-10+18\frac{I}{Q}-6\frac{I^2}{Q^2}+\frac{I^3}{Q^3}-9\frac{I+Q}{I-Q}\right)\ln\frac{I}{Q}.
    \end{split}\label{eq:Ghat}
\end{align}
with 
\begin{equation}
    f(r,t)=\begin{cases} 
      \sqrt{r}\ln \Big|\frac{t-\sqrt{r}}{t+\sqrt{r}}\Big| & \text{for } r>0\\
      0 & \text{for } r=0\\
      2\sqrt{-r}\arctan\frac{\sqrt{-r}}{t} & \text{for } r<0
    \end{cases}.
\end{equation}

\bibliographystyle{JHEP}
\bibliography{ref}

\end{document}